\documentclass[aps,prx,reprint,amsmath,amssymb,amsfonts]{revtex4-1}

\AtBeginDocument{\usepackage{booktabs}}               
\makeatletter
\g@addto@macro\bfseries{\boldmath}
\makeatother

\usepackage[varg]{txfonts}
\usepackage[T1]{fontenc}
\usepackage[utf8]{inputenc}
\usepackage{hyperref}
\usepackage{amsmath}
\usepackage{color}
\usepackage{graphicx}
\usepackage[percent]{overpic}
\usepackage{mathrsfs}
\usepackage{bm}
\usepackage{braket}
\usepackage[nolist,nohyperlinks]{acronym}
\usepackage{comment}

\newcommand{\avg}[1]{\braket{#1}}

\newcommand{\grad}{\vec{\nabla}}
\newcommand{\del}{\partial}

\renewcommand{\vec}[1]{\boldsymbol{#1}}

\newcommand{\vhat}[1]{\vec{\hat{#1}}}

\newcommand{\meV}{\ {\rm meV}}

\newcommand{\s}{\sigma}

\newcommand{\K}{\ {\rm K}}

\newcommand{\J}{J}

\definecolor{cred}{RGB}{228,26,28}
\definecolor{cblue}{RGB}{55,126,184}
\definecolor{cdblue}{RGB}{40,96,139}
\definecolor{clblue}{RGB}{205,223,237}
\definecolor{cgreen}{RGB}{77,175,74}
\definecolor{cgray}{RGB}{150,150,150}
\definecolor{clgray}{RGB}{200,200,200}
\definecolor{cpurple}{RGB}{152,78,163}
\definecolor{corange}{RGB}{255,127,0}
\definecolor{cgold}{RGB}{230,171,2}
\definecolor{cL}{RGB}{255,0,0}
\hypersetup{colorlinks=true,linkcolor=cL,citecolor=cL,urlcolor=cL} 

\begin{document}

\title{Thermal Conductivity of Square Ice}
\author{Ruairidh Sutcliffe} 
\author{Jeffrey G. Rau} 
\affiliation{Department of Physics, University of Windsor, Windsor, Ontario, N9B 3P4, Canada}

\begin{abstract}
We investigate thermal transport in square ice, a two-dimensional analogue of spin ice, exploring the role played by emergent magnetic monopoles in transporting energy. Using kinetic Monte Carlo simulations based on energy preserving extensions of single-spin-flip dynamics, we explicitly compute the (longitudinal) thermal conductivity, $\kappa$, over a broad range of temperatures. We use two methods to determine $\kappa$: a measurement of the energy current between thermal baths at the boundaries, and the Green-Kubo formula, yielding quantitatively consistent values for the thermal conductivity. We interpret these results in terms of transport of energy by diffusion of magnetic monopoles. We relate the thermal diffusivity, $\kappa/C$ where $C$ is the heat capacity, to the diffusion constant of an isolated monopole, showing that the subdiffusive monopole implies $\kappa/C$ vanishes at zero temperature. Finally, we discuss the implications of these results for thermal transport in three-dimensional spin ice, in spin ice materials such as Dy$_2$Ti$_2$O$_7$ and Ho$_2$Ti$_2$O$_7$, and outline some open questions for thermal transport in highly frustrated magnets.
\end{abstract}

\date{\today}

\maketitle


\section{Introduction}
\label{sec:intro}
Highly frustrated interactions in magnetic systems can lead to a rich variety of unusual phases of matter~\cite{lacroix2011introduction}. These include disordered phases, such as spin liquids, that host emergent gauge structures and fractionalized excitations carrying gauge charge~\cite{balents2010spin,castelnovo2012spin,savary2016quantum,gingras2014quantum}. However, identifying magnetic materials that realize such spin liquid states remains a challenge. Detection of fractionalized excitations, either directly via spectroscopic experiments or indirectly through transport experiments, has proven to be a promising route to unambiguously diagnose the presence of a spin liquid phase~\cite{knolle2019field}. 

A paradigmatic example of such magnetic systems are spin ice materials~\cite{spinice}, where the frustration of the magnetic moments mimics the disorder of protons in water ice~\cite{petrenko1999physics,ramirez1999zero,bramwell2001spin}. Here, the magnetic moments lie on the sites of the pyrochlore lattice~\cite{gardner2010magnetic}, a three-dimensional network of corner-sharing tetrahedra. Interactions between the spins are frustrated, enforcing a ``two-in/two-out'' ice rule~\cite{bernal1933theory,pauling1935structure,anderson1956ordering} on each tetrahedron and leading to a macroscopically degenerate ground state manifold~\footnote{In realistic models of spin ice materials, this degeneracy is weak and is lifted by longer range interactions at very low temperatures leading to long-range  order~\cite{melko2004monte}. In the materials themselves, the slowing of the dynamics effectively freezes the system before this transition can take place~\cite{snyder2004low}}. The spin correlations in this ice manifold are dipole-like and realize a version of magnetostatics, a so-called ``Coulomb phase''~\cite{henley2010coulomb,castelnovo2012spin}.

The lowest energy excitations in this Coulomb phase have the spins in a three-in/one-out or three-out/one-in configuration, violating the ice rule. These defects are \emph{fractionalized}, appearing only in pairs, and behaving as emergent magnetic monopoles~\cite{castelnovo2008magnetic,fennell2009magnetic}. The background of the ice rule tetrahedra realize a tensionless tangle of Dirac strings~\cite{jaubert2009signature} linking the monopoles. The entropy of these strings (or, more directly, the magnetostatic dipolar interaction~\cite{castelnovo2008magnetic}) provides a Coulomb interaction between these monopoles, completing the magnetostatic analogy~\cite{henley2010coulomb}. At low temperatures, these monopole excitations are dilute, and spin ice realizes a magnetic analogue of an electrolyte and much of its rich associated physics~\cite{castelnovo2011debye,kaiser2018emergent}. The presence of magnetic monopoles has direct implications for many experimental probes of spin ice, such as magnetic susceptibility and relaxation~\cite{quilliam2011dynamics,yaraskavitch2012spin,bovo2013brownian,kassner2015supercooled,paulsen2016experimental,eyvazov2018common} as well as meso- and nano-scale magnetic noise measurements~\cite{watson2019real,dusad2019magnetic,samarakoon2021anomalous}.

\begin{figure}[tp]
    \centering
    \includegraphics[width=0.75\columnwidth]{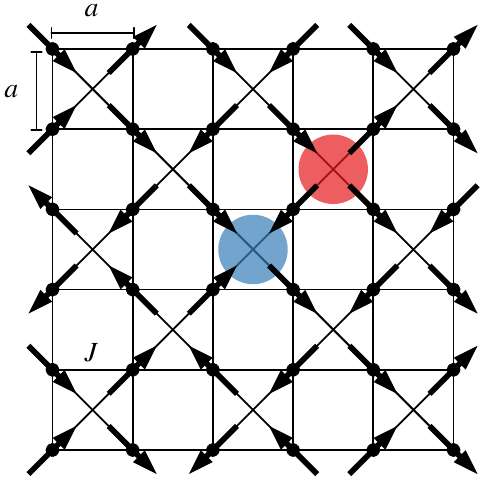}
    \caption{Illustration of the square ice model on the checkerboard lattice, highlighting the lattice spacing, $a$, and nearest neighbour exchange constant, $J$. Superimposed is an (excited) spin state having the majority of tetrahedra satisfying the ice rules with $Q_I=0$ and two hosting monopole defects, one having $Q_I=+1$ and the other $Q_I=-1$ (illustrated by filled circles).}
    \label{fig:lattice}
\end{figure}

A particularly compelling manifestation of mobile magnetic monopoles would be conduction of heat, giving a magnetic contribution to the thermal conductivity at low temperatures, distinct from that of the lattice. While there is a growing body of experimental work exploring thermal transport in frustrated magnets broadly~\cite{hirschberger2015large,kasahara2018majorana,hirschberger2019enhanced}, and in spin ice in particular~\cite{klemke2011thermal,kolland2012thermal,kolland2013anisotropic,toews2013thermal,scharffe2015heat,toews2018}, the interpretation of this data is hindered by the lack of \emph{theoretical} predictions for the thermal conductivity.
Indeed, to date, to the best of our knowledge, there have been only a handful of explicit calculations of thermal transport in classical Ising models~\cite{herrmann1986fast,costa1987energy,harris1988thermal,saito1999transport,agliari2007configurations,casartelli2007heat,masumoto2019diffusion}, with most studies that can access the paramagnetic regime considering only classical Heisenberg models (see, e.g., Refs.~[\onlinecite{blume1974,thermal1,thermal2,thermal3,thermal5,yao2021}]). Going further, for \emph{frustrated} Ising models such as in spin ice, we know of \emph{no} such calculation, even for simple toy models~\footnote{Energy transport in some \emph{disordered} Ising models has been studied~\cite{agliari2009energy,agliari2010microscopic}, though this is distinct from the kind of non-disordered frustration of interest here.}.

In this paper, we explore thermal transport in a minimal frustrated model: \emph{square} ice, a two-dimensional version of spin ice. This lower-dimensional model shares many of the features of the three-dimensional models used to study spin ice materials, such as its macroscopic ground state degeneracy~\footnote{At zero temperature this reduces to the celebrated six-vertex model~\cite{baxter2016exactly}} and fractionalized magnetic monopole excitations. After a brief review of the physics of square ice (Sec.~\ref{sec:ice}) and the phenomenology of thermal transport (Sec.~\ref{sec:dynamics}), we augment the square ice model with energy-conserving Monte Carlo dynamics to access the thermal conductivity (Sec.~\ref{sec:kinetic_monte_carlo}). We compute the thermal conductivity, $\kappa$, using two different techniques (Sec.~\ref{sec:measuring}): using an explicit thermal bath and the Green-Kubo formula, obtaining consistent results. We compare this thermal conductivity to that expected from diffusion of magnetic monopoles, computing the effective diffusion constant $D=\kappa/C$, finding that, at low temperatures, this diffusion constant tends to zero (Sec.~\ref{sec:diffusion}). To understand this result, we directly compute the mean-squared displacement of a monopole at $T=0$, finding that it is subdiffusive, scaling as $\propto t^{\alpha}$ with $\alpha \approx 0.92 < 1$, due to power-law correlations between the steps of its random walk at different times.
Finally, we comment on aspects of these results which we expect to carry over to three-dimensional spin ice, as well as the implications these results may have for experimental studies of spin ice materials such as Dy$_2$Ti$_2$O$_7$ and Ho$_2$Ti$_2$O$_7$ and highlight some open questions.

\section{Square Ice}
\label{sec:ice}

Before delving into our results on thermal transport, we first review some of the important features of the square ice model. The simplest version of square ice is an Ising model defined on a two-dimensional checkerboard lattice with only nearest-neighbour interactions (see Fig.~\ref{fig:lattice}). This can be viewed as a  three-dimensional pyrochlore lattice, as is relevant for spin ice, projected into a two-dimensional plane. We write the energy as
\begin{equation} 
\label{eq:square-ice}
E = J \sum_{\avg{ij}} \sigma_i\sigma_j,
\end{equation}
where $J>0$ is the (anti-ferromagnetic) exchange constant, $\sigma_i = \pm 1$ are Ising spins and $\sum_{\avg{ij}}$ indicates a sum over nearest neighbour bonds of the checkerboard lattice. We refer to the fully connected plaquettes of the checkerboard lattice as ``tetrahedra'', following the three-dimensional nomenclature.

Using this notation, this model can be re-written in a more suggestive form as
\begin{equation}
E = 2J \sum_I Q_I^2 - JN,
\end{equation}
where $I$ denotes a tetrahedron, $N$ is the total number of spins and $Q_I$ is the \emph{charge} defined as
\begin{equation}
\label{eq:monopole-charge}
Q_I \equiv \frac{1}{2}(-1)^I \sum_{i \in I} \sigma_i,
\end{equation}
where $\sum_{i \in I}$ is a sum over the four spins of the tetrahedron $I$.
This charge can take values $Q_I = 0, \pm 1, \pm 2$. The sublattice sign $(-1)^I$ ensures that the total charge is fixed to zero. The ground state manifold of this model is highly degenerate, with any state with all tetrahedra satisfying $Q_I=0$, i.e. two spins positive and two negative,  having a minimal energy $E=-JN$. The high level of degeneracy  manifests itself as a residual entropy at $T=0$, determined, by \citet{lieb1967}, to be exactly~\footnote{Lieb's result assumes frames this constant in terms of the number of water molecules; translating this to spins gives the additional factor of $1/2$.} $S_0 = 3N/4 \log{(4/3)}$. This ground state manifold is equivalent to the (exactly solvable) six-vertex model with equal weight for each vertex type ($a=b=c=1$, $\Delta=1/2$) and thus many of its features can be computed explicitly; we refer the reader to \citet{baxter2016exactly} for details.

Excitations out of this ground state manifold present themselves as tetrahedra with $Q_I = \pm 1$ or $Q_I = \pm 2$. These defects can be interpreted as ``charges'', with $Q_I = \pm 1$ corresponding to (single) monopoles and $Q_I = \pm 2$ to double monopoles. Since $\sum_I Q_I=0$, by definition these charges must appear in oppositely charged pairs to maintain overall neutrality. At low temperatures, $T\ll J$,  the system is predominantly in an ice state with $Q_I=0$ for most $I$, with monopole defects being dilute~\cite{anderson1956ordering,henley2010coulomb}. Na\"ively, we would expect the density of monopoles to go roughly as $\rho_1 \sim e^{-\Delta_1/T}$ where $\Delta_1=2J$ is the energy cost to excite a single monopole. Similarly, the density of double monopoles should go as $\rho_2 \sim e^{-\Delta_2/T}$ where $\Delta_2 = 8J$; this is negligible ($\rho_2 \ll \rho_1$) until temperatures where $T \sim O(J)$ and thus can be typically ignored for $T \ll J$~\cite{castelnovo2011debye}.

\section{Thermal Conductivity}
\label{sec:dynamics}
The thermal conductivity tensor, $\kappa_{\mu\nu}$, relates the difference in temperature across a sample to the energy that flows in response, with
\begin{equation}\label{kappa-general}
J_\mu = -\sum_{\nu}\kappa_{\mu\nu} \nabla_{\nu} T,
\end{equation}
where $\vec{J}$ is the total energy (or heat) current and $\grad T$ is the thermal gradient. To define this energy current more precisely, consider the continuity equation
\begin{equation}
\label{eq:continuity}
\frac{\del \epsilon_i}{\del t} + {\rm div}_i (j) = 0,
\end{equation}
where $\epsilon_i$ is the energy density (satisfying $E = \sum_i \epsilon_i$) and ${\rm div}_i(j) \equiv \sum_j j_{ij}$ is the (lattice) divergence at site $i$ and $j_{ij}$ is the energy current from site $i$ to $j$. This continuity equation ensures that energy is preserved not only globally, but locally as well, so that energy currents are well-defined. 
The total current is then defined as
\begin{equation} \label{eq:total-current}
\vec{\J}=\frac{1}{2}\sum_{ij}(\vec{r}_{j}-\vec{r}_{i})j_{ij},
\end{equation}
where $\vec{r}_i$ is the position of spin $i$. Note that, for the square ice system, lattice symmetries constrain the components of $\kappa_{\mu\nu}$. Invariance under $90^{\circ}$ rotations implies $\kappa_{xx}=\kappa_{yy}$ and $\kappa_{xy}=-\kappa_{yx}$. Absent a magnetic field, time reversal symmetry then requires that $\kappa_{xy}=\kappa_{yx}$ and thus the off-diagonal elements vanish. The thermal conductivity is therefore isotropic, and we define $\kappa_{\mu\nu} \equiv \kappa \delta_{\mu\nu}$.

In order to compute the thermal conductivity, we therefore must have some notion of dynamics that preserves energy locally and globally. Once a specific model and dynamics are chosen, the local energy current $j_{ij}$ can then be determined explicitly through the continuity equation [Eq.~(\ref{eq:continuity})]. We defer discussion of the practicalities of computing $\kappa$ using these definitions to Sec.~\ref{sec:measuring}, first proceeding to define our models and their dynamics.

\section{Kinetic Monte Carlo}
\label{sec:kinetic_monte_carlo}
To enable the calculation of thermal conductivity we extend the square ice model [Eq.~(\ref{eq:square-ice})] to a \emph{kinetic} Ising model that includes discrete time steps and locally preserves the energy.  
To do this, we broadly follow the strategy of interpreting the discrete-time dynamics of a Monte Carlo simulation as a proxy for the true dynamics of an Ising system, as has been successfully used to understand the dynamics of three-dimensional spin ice~\cite{castelnovo2010thermal,jaubert2009signature}. In this class of methods, one sweep through the lattice (one proposed update per spin) is defined as taking place over a time-step of duration $\delta t$.

The simplest type of updates that can be used in Monte Carlo simulations are those which only change the spins at a single site, so-called ``single-spin-flip'' dynamics.
Using this method, a flip of a single spin of the system is proposed at random, say site $k$, resulting in a change in energy of
\begin{equation} \label{eq:spin-flip-energy}
\Delta E_k =-2J\sigma_k \sum_{\avg{ik}}\sigma_i.
\end{equation} 
In a conventional simulation in the canonical ensemble (say using a Metropolis update) at temperature $T$, such a flip is accepted with probability ${\rm min}(1,e^{-\Delta E_k/T})$, working with units where $k_{\rm B}=1$. However, updates governed by such dynamics are not necessarily energy-preserving. To ensure these single-spin-flip updates locally preserve energy, as is necessary to examine thermal conductivity, two modifications to this model are introduced. First, the single-spin-flip method is augmented with a local ``demon'' bath~\cite{creutz1983,herrmann1986fast,harris1988thermal,harris1990demons} to maintain a fixed total energy. Second, a closely related micro-canonical single-spin-flip method~\cite{creutz1986deterministic,herrmann1986fast,costa1987energy} is considered that does not introduce any new degrees of freedom.~\footnote{We expect the use of, say, Glauber dynamics~\cite{glauber1963time} instead of Metropolis dynamics in either method to produce qualitatively similar results for the thermal conductivity, though quantitative details may change.} 

\subsection{Single-Spin-Flip with Demons}
\label{sec:dynamics:ssf-demons}
The first method which we use to ensure energy preservation in the system introduces auxiliary variables that exchange energy with the spins. These variables give or receive the required energy change $\Delta E_k$ of each spin-flip and thus keep the total energy of the system unchanged.

The notion of including extra degrees of freedom, so-called ``demons'' (following Maxwell~\cite{maxwellsdemon}), as a method for performing micro-canonical simulations in statistical physics was first introduced by \citet{creutz1983}. Effectively, it serves as a simple, explicit model for a thermal bath; by allowing the system of interest to exchange energy with the demons, the spins and the demons can reach a mutual (thermal) equilibrium~\cite{creutz1983,harris1990demons}.

Explicitly, $N$ independent demons are introduced, one for each site of the system. Each demon acts locally, interacting only with the spin at the same site and can store an arbitrary \emph{positive} amount of energy, $D_i \geq 0$. The total energy of the system, $E$, is the sum of the energy of the spins [Eq.~(\ref{eq:square-ice})] and the energy of the demons
$$
E = J \sum_{\avg{ij}} \sigma_i\sigma_j +
\sum_i D_i \equiv E_{\rm spin} + E_{\rm demon}.
$$
To ensure energy conservation, for each proposed spin-flip, the update is accepted or rejected based on whether the demon can provide the required change in energy.

Each proposed spin-flip proceeds as follows: if $\Delta E_k \leq 0$ then the demon simply absorbs that energy and the move is accepted with $\sigma_k \rightarrow -\sigma_k$ and $D_k \rightarrow D_k-\Delta E_k$. However, if $\Delta E_k \geq 0$, then the update is only accepted if $D_k \geq \Delta E_k$ since $D_k$ must remain positive. In this way, although energy is not conserved for the spins alone, it \emph{is} conserved by the combination of the spins and the demons, with each update maintaining a constant $E_{\rm spin} + E_{\rm demon}$~\cite{creutz1983}. 

In contrast to the traditional Metropolis update algorithm, which simulates within the canonical ensemble at some fixed predefined temperature, with the addition of demons, the total energy is conserved and the updates no longer depend on an explicit temperature parameter~\cite{creutz1983,harris1990demons}. However, at late times, we expect that the demons and the spins will reach a (mutual) thermal equilibrium with a well-defined temperature, $T$.
Following \citet{creutz1983}, this temperature can be determined using the expectation values of the demon energies and thus a (local) temperature can be defined for each demon. Since the demons are in thermal contact with the spins, at late times when thermal equilibrium between the two is reached, their temperatures must be equal.

To use the demons as local thermometers,  the local temperature must be related to some observable associated with the demon. Noting that the energy changes $\Delta E_k$ come only in multiples of $\pm 4J$, the energies of the demon can thus take only values of the form $D_k = 4J n_k$ where $n_k=0,1,2,\cdots,\infty$. Therefore, in equilibrium, the statistical mechanics of each demon is equivalent to a quantum harmonic oscillator with energy spacing $4J$. Specifically, in thermal equilibrium, this implies that the average demon energy at site $k$, $\avg{D_k}$, is given by
\begin{equation} \label{eq:demon-energy}
\avg{D_k}=\frac{4J}{e^{4J/{T_{k}}}-1},
\end{equation}
where $T_k$ is the (local) temperature of the demon.
This can be inverted to obtain
\begin{equation} \label{eq:demon-temperature}
T_k=\frac{4J}{\log\left(1+\frac{4J}{\avg{D_k}}\right)}.
\end{equation}
When the system has reached equilibrium we would expect all demons to have reached a common temperature, with $T_k \equiv T$; this is not necessarily true when the system is not uniform or the system has not yet reached equilibrium.

\subsection{Micro-Canonical Single-Spin-Flip}
\label{sec:dynamics:micro-canonical-ssf}

As the true dynamics of the Ising spins, say in spin ice materials, is certainly more complicated than the ad-hoc model using demons defined here, it will be useful to compare these results to other choices of dynamics to clarify what features are strongly dependent on this choice.

To this end, we also consider a micro-canonical variant of the single-spin-flip update. As in Sec.~\ref{sec:dynamics:ssf-demons}, a spin $k$ is chosen at random and again the change in energy required to flip this spin, $\Delta E_k$, is calculated via Eq.~(\ref{eq:spin-flip-energy}). However, unlike the demon method, this spin-flip is only accepted if $\Delta E_k=0$. Similar dynamics have been studied as an example of a cellular automaton~\cite{vichniac1984simulating}, and in modelling thermal transport in ferromagnetic Ising models~\cite{saito1999transport}. Since this method is equivalent to the demon method provided $\Delta E_k$ is forced to be zero, the expression for energy current will be identical if we set $\Delta E_k=0$ [see Eq.~(\ref{eq:local-energy-current})]. It should be emphasized, however, that without the demons it is difficult to determine the local temperature of the system. This affects the available methods for computing $\kappa$; while we can use both the thermal bath method (Sec.~\ref{sec:thermal-bath}) and the Green-Kubo method (Sec.~\ref{sec:green-kubo}) for demon dynamics, for the micro-canonical single-spin-flip dynamics we can only use the Green-Kubo approach.

\section{Measuring Thermal Conductivity}
\label{sec:measuring}
Having outlined the kinetic Monte Carlo methods used, we now discuss how the thermal conductivity, $\kappa$, is determined. First, the system is augmented with thermal baths at different temperatures at opposing ends, between which the induced energy current can be measured and thus $\kappa$ determined. Secondly, $\kappa$ is computed directly from the Green-Kubo~\cite{kubo1957,green1954markoff} formula. This method does not rely on the baths, but does require knowledge of the total energy current in terms of the spins.

\subsection{Thermal Bath Method}
\label{sec:thermal-bath}
\begin{figure}[tp]
    \centering
    \includegraphics[width=\columnwidth]{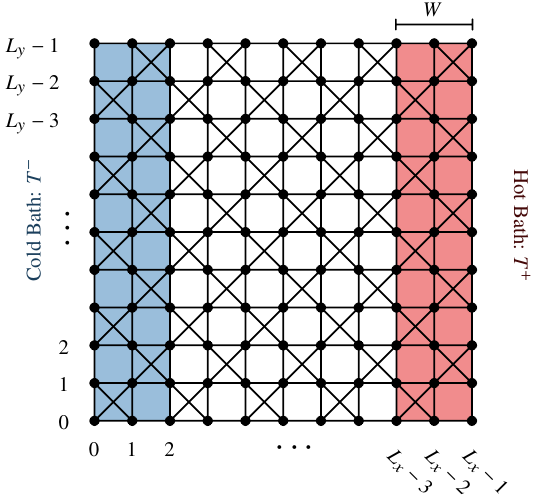}
    \caption{Illustration of a checkerboard lattice of dimensions $L_x\times L_y$ with thermal bath regions of width $W=3$ and temperatures $T^{+}$ (the hot bath) and $T^{-}$ (the cold bath). Open boundary conditions are imposed along the $\vhat{x}$ direction with the periodic boundary conditions imposed along $\vhat{y}$. The thermal bath induces a temperature gradient along $\vhat{x}$, with energy current, $J$, flowing from the hot bath to the cold bath. Due to translational symmetry along $\vhat{y}$ and conservation of energy this current is uniform.
    }
    \label{fig:bath-diagram}
\end{figure}
Conceptually, the simplest strategy to determine $\kappa$ is to (loosely) emulate the experimental protocol: by holding the ends of a finite sample at a (small) temperature difference $\Delta T$, an energy current, $J^x$, is induced with $\kappa = J^x/\Delta T$.
The presence of explicit baths allows the total current flowing from one bath to the other to be computed straightforwardly, while the local thermometers provided by the demons allows access to the induced temperature gradient. 

Explicitly, consider a finite sample geometry, as shown in Fig.~\ref{fig:bath-diagram}. Open boundary conditions are imposed along the $\vhat{x}$ direction, defining two ``bath'' regions of width $W$ sites on the left and right edges of the sample. In the $\vhat{y}$ direction, the usual periodic boundary conditions are kept. To keep the bath regions at some fixed temperature, spin-flips in those regions are performed using canonical Metropolis updates. The remainder of the system is updated using the energy preserving demon dynamics (Sec.~\ref{sec:kinetic_monte_carlo}).

In this setup, one bath is held at a low temperature, $T^-$, with the other being held at a higher temperature, $T^+$, inducing a site-dependent temperature $T_i$ across the sample. Due to translation symmetry along $\vhat{y}$, $T_i$ can be written as $T_i \equiv T(x_i)$, depending only on the coordinate along $\vhat{x}$. For small temperature differences, $T^+-T^-$, the variation of $T(x)$ should be linear in $x$, while for larger changes it can become non-linear.
The temperature at fixed $x$ can be determined by averaging the demon energy along the $\vhat{y}$ direction, $\frac{1}{L_y}\sum_{i_y}\avg{D_{i_x,i_y}}$ for each $i_x$ and then computing the temperature as in Eq.~(\ref{eq:demon-temperature}).

The total current flowing between the baths, $J^x$, can be computed directly from the energy imparted to the system in the bath regions by the Metropolis updates. Since open boundary conditions have been implemented along the $\vhat{x}$ direction, and the bulk (non-bath region) preserves energy, energy can only enter and leave the system through the two baths.
Translation along $\vhat{y}$ then implies that the total current passing through a surface with fixed $x$ is \emph{constant}, independent of $x$, throughout the bulk of the system.
More concretely, the net energy flow into the cold bath, say $\J^x_{\mathcal{B}_-}$, must be equal and opposite to the net flow into the opposite bath, $\J^x_{\mathcal{B}_+} = -J^x_{\mathcal{B}_-}$. We thus can identify the current flowing through the middle of the sample:
$$
J^x \equiv J^x_{\mathcal{B}_+} \equiv \frac{\J^x_{\mathcal{B}_+}-\J^x_{\mathcal{B}_-}}{2}
$$
where the latter expression provides improved statistics.
The energy flow into and out of the baths per time step (that is, per sweep) can be computed directly from the energy changes $\Delta E_i$ of the Metropolis updates. For an $L_x$ by $L_y$ system, we define
\begin{equation}
J^x_{\mathcal{B}_{\pm}} \equiv \frac{1}{L_y}\sum_{i \in \mathcal{B}_{\pm}}\avg{\Delta E_i},
\end{equation}
where $\sum_{i \in \pm}$ denotes a sum over the site belonging to the left ($\mathcal{B}_-$, blue region in Fig.~\ref{fig:bath-diagram}) or right ($\mathcal{B}_+$, red region in Fig.~\ref{fig:bath-diagram}) baths. Note that once the system has reached equilibrium, each bath current, $J^x_{\mathcal{B}_{\pm}}$, becomes independent of Monte Carlo time.

Once $J^x$ and $T_i$ are computed, $\kappa$ can be directly determined from its definition. For a one-dimensional geometry with isotropic thermal conductivity and slowly varying temperature $T(x)$, Eq.~(\ref{kappa-general}) reduces to
$$
J^x = -\kappa(T(x)) \frac{dT(x)}{dx}.
$$
Since the energy current, $J^x$, is a (known) constant, this can be written
\begin{equation}
\label{eq:kappa4}
\kappa(T(x))=-J^x \left(\frac{dT(x)}{dx}\right)^{-1},
\end{equation}
giving $\kappa$ at temperature $T(x)$ from the derivative $dT/dx$.

Absent corrections from non-linear thermal conductivities, this relation holds for \emph{arbitrary} temperature differences $\Delta T \equiv T^+- T^-$ between the baths. Due to inversion symmetry, the next term is expected to appear at third-order, with
$$
J^x = -\kappa \frac{dT}{dx} - \kappa_3 \frac{d^3 T}{dx^3} + \cdots
$$
so then as long as $\kappa_3 (d^3T/dx^3) \ll \kappa (dT/dx)$ then $\kappa(T(x))$ can be extracted.
Therefore, if $T(x)$ is changing slowly enough, we do not need to restrict $T^{\pm}$ to the small temperature differences necessary to induce a linear temperature gradient. Indeed, by choosing $T^{-} \approx 0$ on a sufficiently wide lattice (to suppress higher derivatives) $\kappa(T)$ can be obtained over the full range $0 \lesssim T \lesssim T_+$ from a single simulation. 
Practically, on the lattice we approximate the derivative in Eq.(\ref{eq:kappa4}) using central differences to obtain
\begin{equation}
\kappa(T_i)=-\frac{2J^x}{T_{i+x}-T_{i-x}},
\end{equation}
where $T_{i\pm x}$ are the temperatures of the columns to left and right of the column $i$ and the lattice spacing, $a$, is set to be unity throughout.

Note that we have confirmed that the thermal conductivity obtained for sufficiently large lattice sizes (along the gradient direction) and different bath temperature differences $\Delta T$ yield consistent values for $\kappa$. We have also confirmed that these results are independent of the sizes of the baths, $W$, so long as the bath width is larger than one.

\begin{figure}[tp]
    \centering
    \includegraphics[width=\columnwidth]{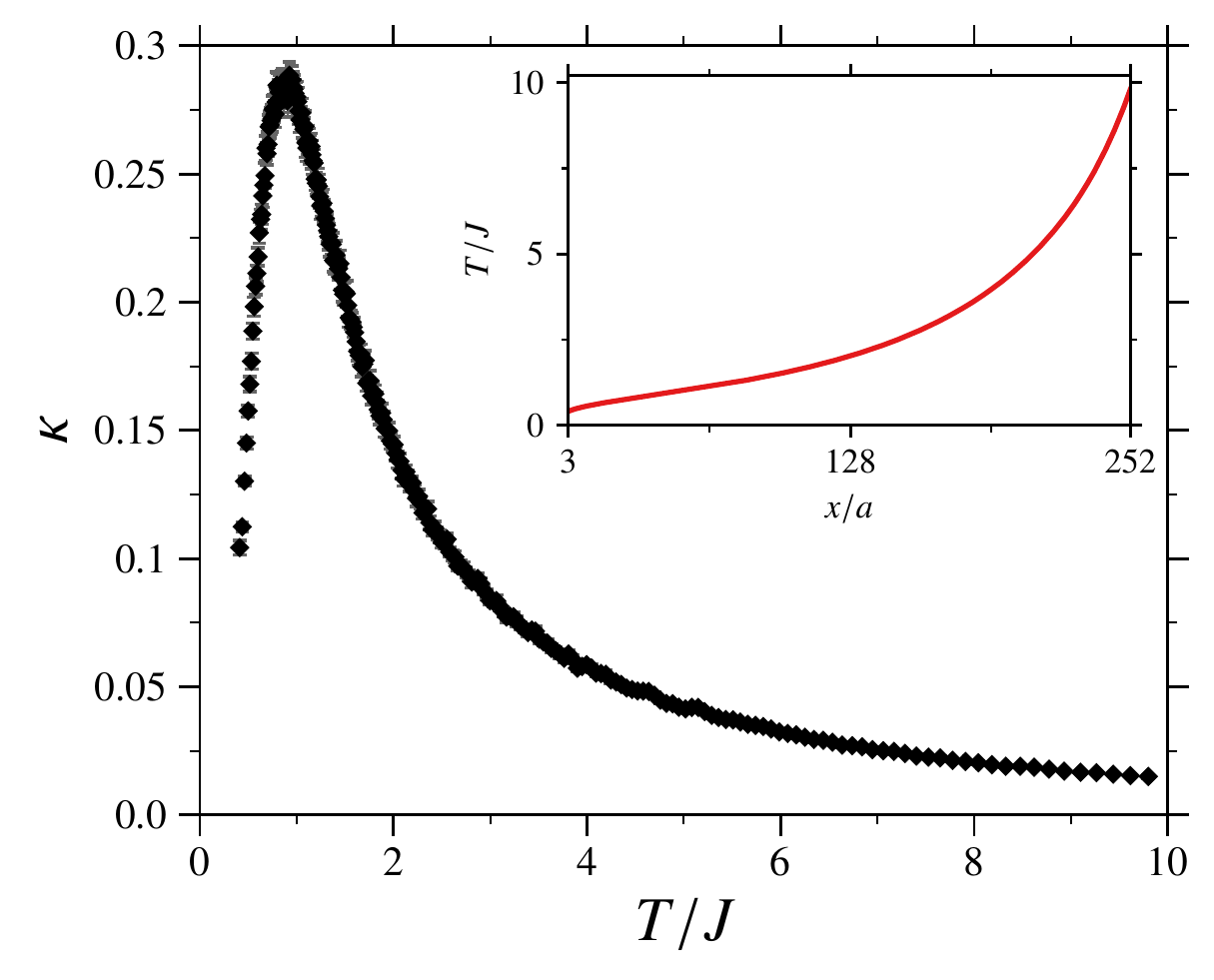}
    \caption{Thermal conductivity, $\kappa$, for a $256\times 128$ system computed using the thermal bath method (see Sec.~\ref{sec:thermal-bath}) with bath size $W=3$ as a function of temperature, $T/J$. Natural units with $J$, $a$, $\delta t$ and $k_{\rm B}$ set to unity are used. The inset shows the temperature gradient induced by the baths as a function of position along $\vhat{x}$, calculated via Eq.~(\ref{eq:demon-temperature}) using the demon energies averaged along the $\vhat{y}$ direction, with bath temperatures $T^-/J=0.1$ and $T^+/J=10.0$.
    Error bars shown are estimated using a standard bootstrap method~\cite{newman1999monte}.
}
     \label{fig:baths-w-gradient}

\end{figure}

The thermal conductivity, $\kappa$, calculated using the thermal bath method is shown in Fig.~\ref{fig:baths-w-gradient}. 
At high temperature, $T \gg J$ transport is incoherent, reflecting fluctuations in the energy current with $\kappa \propto \avg{J^2_x}/T^2$ (also see inset Fig.~\ref{fig:kubo-v-micro}), as expected~\cite{harris1988thermal,saito1999transport}. For temperatures $T\sim J$, $\kappa$ increases to a maximum at $T/J \approx 1$ before dropping quickly toward zero for $T \ll J$. This is in line with expectations from transport by monopole excitations. At low temperatures the monopoles are dilute with exponentially small density $\propto e^{-2J/T}$ and thus $\kappa$ is small. As the temperature is raised, the increased population of monopoles raises $\kappa$ until $T \sim J$ where the description in terms of a dilute monopole gas begins to break down. 

The inset of Fig.~\ref{fig:baths-w-gradient} shows the variation of the local temperature as a function of position $x/a$. Due to the large temperature difference between the two baths ($T^+/J=10.0$ and $T^-/J=0.1$) the local temperature $T(x)$ is \emph{not} a simple linear function of $x$, but has significant curvature. However, since the current is constant, and the system is sufficiently wide, $\kappa(T)$ can still be reliably extracted, though the computed points become sparser at low temperatures.

\subsection{Green-Kubo Formula}
\label{sec:green-kubo}
To further confirm that the non-linear temperature distribution involved in the thermal bath method yields the correct thermal conductivity, we also consider a direct calculation of $\kappa$ using the Green-Kubo formula (see App.~\ref{app:green-kubo} for a relevant derivation). In contrast to the bath method, to use this formula, a precise definition of the local current $j_{ij}$ [Eq.~(\ref{eq:continuity})] that defines the total current $\vec{J}$ [Eq.~(\ref{eq:total-current})] is needed.

The local energy current $j_{ij}(t)$ from site $i$ to site $j$, during a specified sweep at time $t$, can be written as the sum over currents induced by each spin-flip. Explicitly, the current from a (proposed) flip at site $k$ is written as $j^{(k)}_{ij}$ with (see App.~\ref{app:energy-current} for a derivation)
\begin{equation}\label{eq:local-energy-current}
j_{ij}^{(k)}= \begin{cases}
J\sigma_{i}\sigma_{j}\left( \delta_{ik}-\delta_{jk}\right),  &D_k \geq \Delta E_k, \\
0 & D_k < \Delta E_k,
\end{cases}
\end{equation}
where $J$ is the exchange constant (not to be confused with the total current) and $\delta_{ik}-\delta_{jk}$ governs the direction of the current flow, ensuring $j_{ij}^{(k)}=-j_{ji}^{(k)}$. The energy current, $j_{ij}(t)$, can then be computed by adding up the $j_{ij}^{(k)}$ over the sweep, updating the spin configuration after each step. For the micro-canonical single-spin-flip case, the expression is identical, except for the restriction that $\Delta E_k=0$. 

From the $j_{ij}(t)$ time series for the total current, $\vec{J}_t$ [Eq.~(\ref{eq:total-current})] can be obtained, which can then be used to determine $\kappa$ via the Green-Kubo formula~\cite{kubo1957}
\begin{equation} \label{eq:Kubo}
\kappa=\frac{1}{N T^2}\sum_{\tau=0}^{\infty}\langle J_{\tau}^{x}J_{0}^{x}\rangle(1-\frac{1}{2}\delta_{\tau,0}),
\end{equation}
where $N$ is the number of spins and $T$ is the temperature. Note that we continue to use natural units where the lattice spacing, $a$, time-step $\delta t$, and $k_{\rm B}$ are set to unity~\footnote{
Restoring these factors adds a factor of ${\delta t}/{k_B}$ to the two-dimensional thermal conductivity giving units of $W/K$. For the three-dimensional thermal conductivity with units $W/(K m)$, one must also divide by the layer spacing.
}.

\begin{figure}[tp]
    \centering
    \includegraphics[width=\columnwidth]{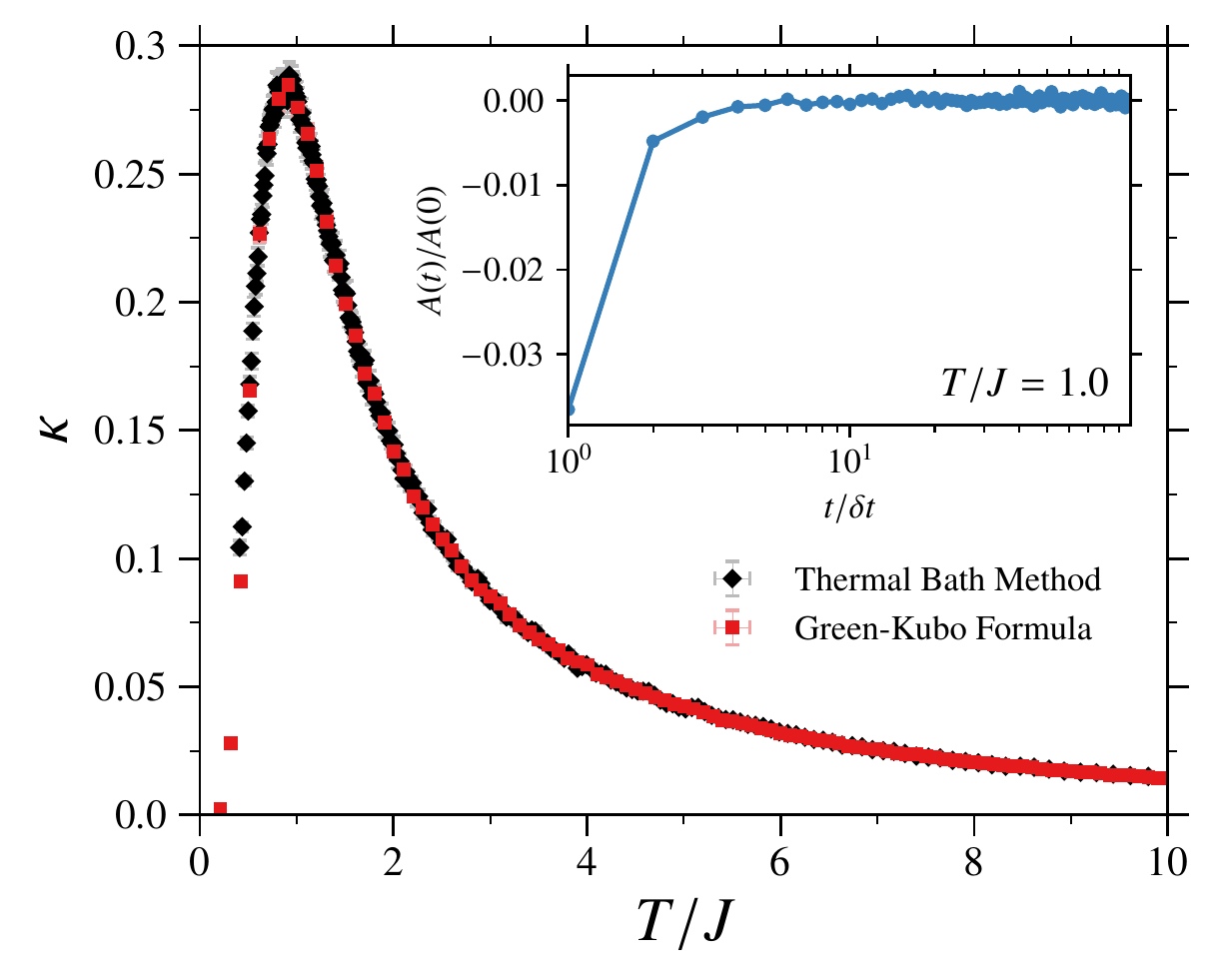}
    \caption{Comparison of the thermal conductivity, $\kappa$, using demon dynamics as a function of temperature for a $256\times 128$ system computed using the thermal bath method (Fig.~\ref{fig:baths-w-gradient},  Sec.~\ref{sec:thermal-bath}) and for a $256\times 256$ system using computed using the Green-Kubo formula (Sec.~\ref{sec:green-kubo}) , showing excellent agreement. The inset shows the relevant energy current auto-correlation function (at $T/J=1.0$) that appears the Green-Kubo formula, Eq.~(\ref{eq:Kubo}), scaled by the dominant $t=0$ element, showing that it quickly decays to zero.
    }    
    \label{fig:kubo-v-baths}    
\end{figure}

This formula can be viewed as the auto-correlation function for the energy current, $A(\tau)$, defined as (noting that $\avg{\vec{J}_\tau}=0$)
\begin{equation}
A(\tau) \equiv \avg{J^x_{\tau} J^x_0} - \avg{J^x_{\tau}}\avg{J^x_0},
\end{equation}
integrated over all times and scaled by $1/{NT^2}$, with the $\tau=0$ component having half weight. Practically, when computing this auto-correlation function, noise dominates at large values of $\tau$. We thus do not extend the sum over the total number of sweeps, but instead exploit that at large times $A(\tau)$ should decay quickly (see Fig.~\ref{fig:kubo-v-baths}). The sum can thus be truncated at some time much smaller than the total number of sweeps; the range $0 \leq \tau \leq 10^2$ was typically used in our simulations. We have confirmed that our results are not sensitive to the precise choice of this cutoff.

As Eq.~(\ref{eq:Kubo}) makes use of the overall temperature, $T$, we must ensure that the system can be initialized at, or at least near, a desired temperature. To do this, the initial total energy $E$ (which is conserved by the dynamics) is chosen so that it matches the expected energy, $\avg{E}$, found in the canonical ensemble at the desired temperature, $T$. How this is done depends on the dynamics chosen; different initialization methods are used for the demon dynamics vs. the micro-canonical single-spin flip dynamics.

For the method with demons (Sec.~\ref{sec:dynamics:ssf-demons}), we employ a strategy to approximately reach a desired temperature, without needing to perform additional Monte Carlo simulations to initialize the state. First, the spins are set to be in a ground state, meaning all tetrahedra having $Q_I=0$ [Eq.~(\ref{eq:monopole-charge})] giving $E_{\rm spin}=-JN$. The total demon energy is chosen such that the combination of the spins and the demons gives approximately the desired temperature. To do this, the total energy in the canonical ensemble is approximated as
$$
E(T) \approx N \left(E_1(T) +\frac{4J}{e^{4J/{T}}-1}\right),
$$
where $E_1(T)$ is the energy per spin in the single tetrahedron approximation (see, e.g., Ref.~\cite{singh2012corrections}), where only four spins are kept. Explicitly, 
\begin{equation}\label{eq:single-tet-energy}
E_1(T)=\frac{6J(e^{6J/T}-e^{2J/T})}{Z_1(T)},
\end{equation}
where $Z_1(T)\equiv 2e^{-6J/T}+6e^{2J/T}+8$ is the partition function for the single tetrahedron. If we choose the initial demon energy to then be
\begin{equation}
E_{\rm demon} = N \left(E_1(T) +\frac{4JN}{e^{4J/{T}}-1} + J \right),
\end{equation}
then $E_{\rm spin} + E_{\rm demon} \approx E(T)$. Practically, the demon energies are randomly incremented until the total demon energy reaches this value.
Importantly, the final temperature of the spins used in Eq.~(\ref{eq:Kubo}) is computed using the average demon energy and \emph{not} this initial target temperature; the use of the single-tetrahedron approximation thus only serves to approximately obtain the desired set of temperatures.

Note that without the demons we do not have a convenient way to impart the initial state with a given energy $E$. For the micro-canonical single-spin-flip simulations (Sec.~\ref{sec:dynamics:micro-canonical-ssf}) we therefore fall back to performing a Monte Carlo simulation using a standard Metropolis algorithm at temperature $T$ with single-spin-flip updates (until thermalization) to generate an initial state with the desired energy (see Sec.~\ref{sec:kinetic_monte_carlo}).

The results for $\kappa$ computed using the Green-Kubo formula are shown in Fig.~\ref{fig:kubo-v-baths}, along with the $\kappa$ computed using the bath method.
One can see that they agree \emph{quantitatively}, confirming our assertion that the non-linear temperature gradients involved in the thermal bath method do not affect the results. The same features noted in Sec.~\ref{sec:thermal-bath} can be seen, with essentially perfect agreement between the results from the thermal bath and Green-Kubo methods. The inset of Fig.~\ref{fig:kubo-v-baths} shows the energy current auto-correlation function that appears in the Green-Kubo formula [Eq.~(\ref{eq:Kubo})]. One can see that it decays very quickly; by $t\sim O(10^1)$ it is essentially zero, thus a cutoff larger than this has little effect on the thermal conductivity. Note that $A(t)$ in the inset is scaled by the dominant $t=0$ element, with the next largest contribution being $\lesssim 5\%$  of  $A(0)$. The dominance of $A(0)$ persists even to low temperature with $T/J\leq 1.0$. The approximation $\kappa\approx\avg{J_x^2}/(2 N T^2)$ is thus surprisingly accurate over the whole temperature range. This quick decay of $A(t)$ does not depend strongly on temperature; though we only showed $A(t)$ for $T/J=1.0$ in the inset of Fig.~\ref{fig:kubo-v-baths}, these features do not change qualitatively for larger or smaller values of $T/J$.~\footnote{More precisely, $A(1)\sim -0.03A(0)$ is ubiquitous across the full temperature range, while the next element, $A(2)$ shows some temperature dependence; for $T \lesssim J$ having more weight, $A(2)/A(0)\sim -10^{-2}$, while for $T \gtrsim J$ one has $A(2)/A(0)\sim -10^{-3}$.}

Although both the thermal bath method and Green-Kubo formula produce results for $\kappa$ that agree quantitatively, there are some notable practical differences between the two methods. First, the thermal bath method is significantly more efficient with respect to computation time, as one simulation, with a run time which scales as $O(N)$, yields the thermal conductivity for $O(L)$ temperature points. Conversely, when using the Green-Kubo formula directly, a single simulation, with run time still scaling as $O(N)$, produces the value of thermal conductivity  for only a single temperature. Conversely, the thermal bath method has the additional complication that one must ensure that the temperature gradient, $dT/dx$, remains sufficiently small, which is dependent on the size of the (apriori unknown) non-linear corrections. Thus, in order to produce reliable results using the thermal bath method, one must choose the temperatures, $T^{\pm}$, at which to hold the baths and the length of the system along the gradient such that these corrections are negligible. Note that using the Green-Kubo method also requires initialization of the demon energies to fix the system at a specific temperature; when baths are present the temperature is fixed naturally.

\subsection{Dependence on Choice of Dynamics}

\begin{figure}[tp]
    \centering
    \includegraphics[width=\columnwidth]{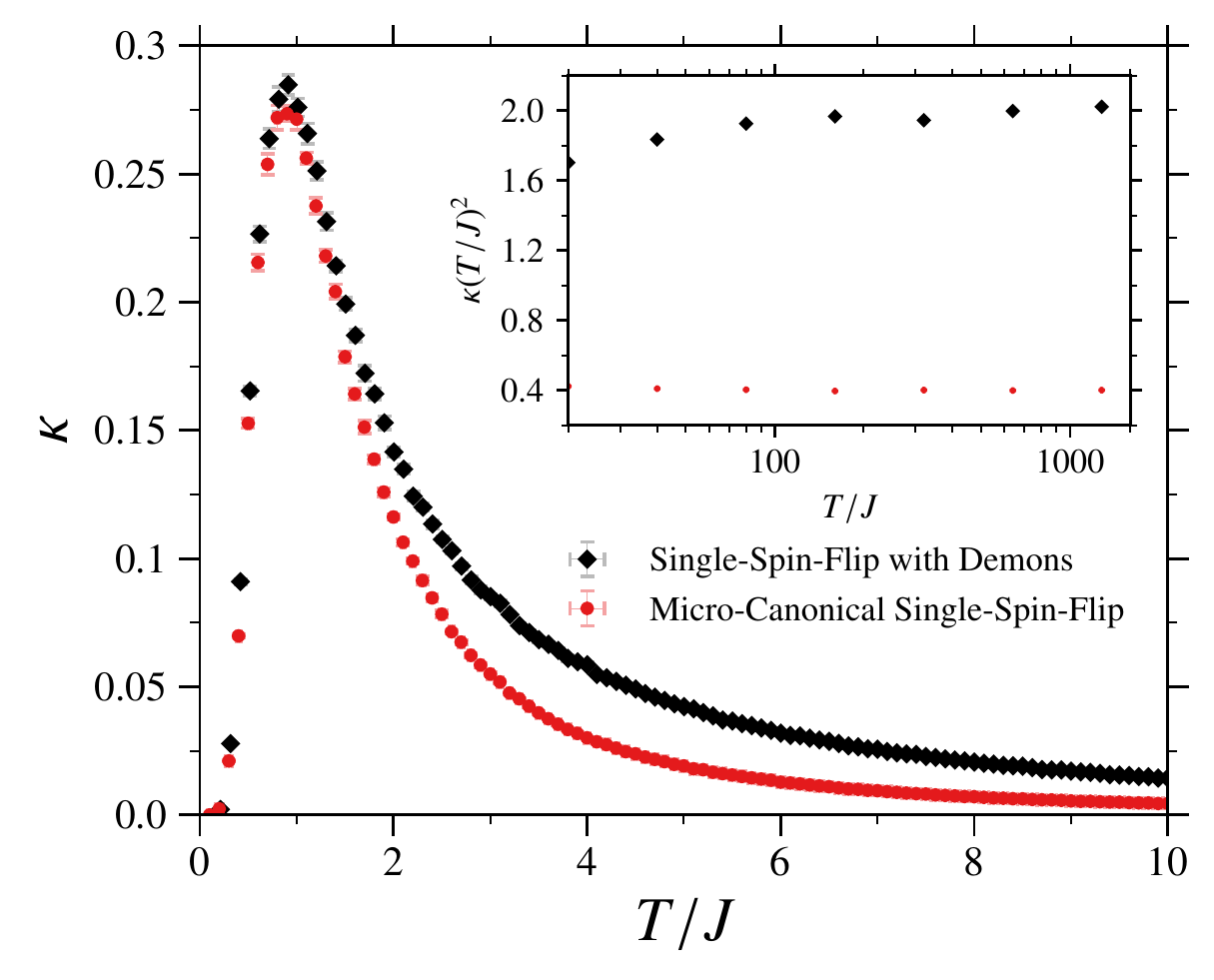}
    \caption{Comparison of the thermal conductivity, $\kappa$, as a function of temperature for a $256\times 256$ system using the single spin-flip method with demons and the micro-canonical single spin-flip method, both computed using the Green-Kubo formula. Agreement is quantitative at low temperature, but only qualitative at high temperatures. Since at high temperature $\kappa \propto 1/T^2$, the inset shows $\kappa T^2$ for the two methods. This $\kappa T^2$ value can be related to the fluctuations of the energy current, $\avg{J_x^2}$, as discussed in the text. }
    \label{fig:kubo-v-micro}
\end{figure}

In order to examine the dependence of these results on the details of the dynamics,  we computed $\kappa$ using the alternate micro-canonical dynamics, as described in Sec.~\ref{sec:dynamics:micro-canonical-ssf}. A comparison of the thermal conductivity for these two choices is shown in Fig.~\ref{fig:kubo-v-micro}. There is quantitative agreement between the two methods at low temperature, with both showing $\kappa\rightarrow 0$ as $T\rightarrow 0$ and maxima of similar magnitude at $T\sim O(J)$. However, at high temperature ($T\gtrsim J$) the two methods agree qualitatively, with both methods going as $\propto (J/T)^2$, but disagree quantitatively with the coefficients of the $O((J/T)^2)$ being different. The inset of Fig.~\ref{fig:kubo-v-micro} highlights this behaviour, showing $(T/J)^2 \kappa$ in the high $T$ regime. Both approach constants, with the micro-canonical method scaling as $\sim 0.4(J/T)^2$ while the demon method scaling as $\sim 2(J/T)^2$ over the same temperature range.  For the demon method, this coefficient of $2$ can be reproduced via a high-temperature expansion, assuming that $\kappa T^2 \propto \avg{J_x^2}$ and currents induced during each sweep are uncorrelated.

Much of this difference in the two methods arises from differences in the likelihood of flipping a spin at high $T$. Specifically, for $T\gg J$, if the spin-flips are uncorrelated, one can approximate $\kappa T^2 \sim \avg{J_x^2}/(2N)$ [Eq.~(\ref{eq:Kubo})], with the scale of the current fluctuations being directly related to the acceptance rate. Using the demon method, at high $T$, every spin-flip is accepted as the demons will always have enough energy to flip the spin irrespective of the cost. In contrast, when using the micro-canonical method, a spin $k$ is only flipped if $\Delta E_k=0$. At high temperature, where all possible spin configurations of the neighbours are equally likely, this amounts to a $5/16 \sim 31.3\%$ probability of a spin-flip. We would na\"ively expect, at least to a first approximation, provided we have uncorrelated spin-flips in both cases, that the ratio of thermal conductivities to have this ratio for $T \gg J$. However, as shown in the inset of Fig.~\ref{fig:kubo-v-baths}, we see a ratio closer to $\sim 20\%$. This small remaining discrepancy is likely due to the fact that spin-flips are significantly more correlated when using the micro-canonical dynamics, with the $t>0$ elements of the current auto-correlation function $A(t)$ remaining important even at high temperature.

\section{Diffusion of monopoles}
\label{sec:diffusion}

It is natural to interpret the thermal conductivity as arising from the motion of the monopoles transporting energy. Since the monopoles are excitations out of the ground state manifold with $Q_I = \pm 1$ they carry a finite amount of energy $\Delta_1=2J$. Since these monopole defects are not fixed and can move through the system at no energetic cost as the spins on the monopole tetrahedra are flipped, they can serve as vehicles for the transport of energy within the system.

We can see this explicitly from a ``macroscopic'' perspective, via the continuity equation for energy density
$$
\frac{\del\epsilon}{\del t}=-\vec{\nabla}\cdot\vec{j},
$$
where $\vec{j}(\vec{r},t)=-\kappa\vec{\nabla}T(\vec{r},t)$ is the local energy current and we assume both $\epsilon$ and $T$ vary slowly with position and time.  Assuming each region of the system is in a local thermal equilibrium, $\epsilon$ and $T$ can be related through $\epsilon(\vec{r},t)\equiv\epsilon(T(\vec{r},t))$. This implies that
$$
\vec{\nabla}\epsilon=\left(\frac{\del\epsilon}{\del T}\right)\vec{\nabla}T,
$$
where $C\equiv \del\epsilon/\del T$ is the heat capacity per spin (equivalent to per volume in natural units). The energy density thus obeys a diffusion equation
\begin{equation}
\frac{\del\epsilon}{\del t}\approx D \nabla^2\epsilon,
\end{equation}
where $D \equiv \kappa/C$ is the thermal diffusivity. At low temperature where the density of defects is low, the energy density can be related to the monopole density, as
$$
\epsilon(\vec{r},t) \approx \epsilon_0 + \Delta_1 n_1(\vec{r},t)+\cdots,
$$
where $\epsilon_0$ is the energy density of the ice background and $n_1$ is the monopole density. A similar diffusion equation for $n_1$ then follows, with
\begin{equation}
\frac{\del n_1}{\del t}\approx D \nabla^2 n_1 ,
\end{equation}
Thus, at low temperature the monopole diffusion constant, $D$, is identical to the thermal diffusivity, $\kappa/C$. This result can also be obtained directly from the Green-Kubo formula for $\kappa$, though through a more complicated argument (see App.~\ref{app:green-kubo-diffusion}).

\begin{figure}[tp]
    \centering
    \includegraphics[width=\columnwidth]{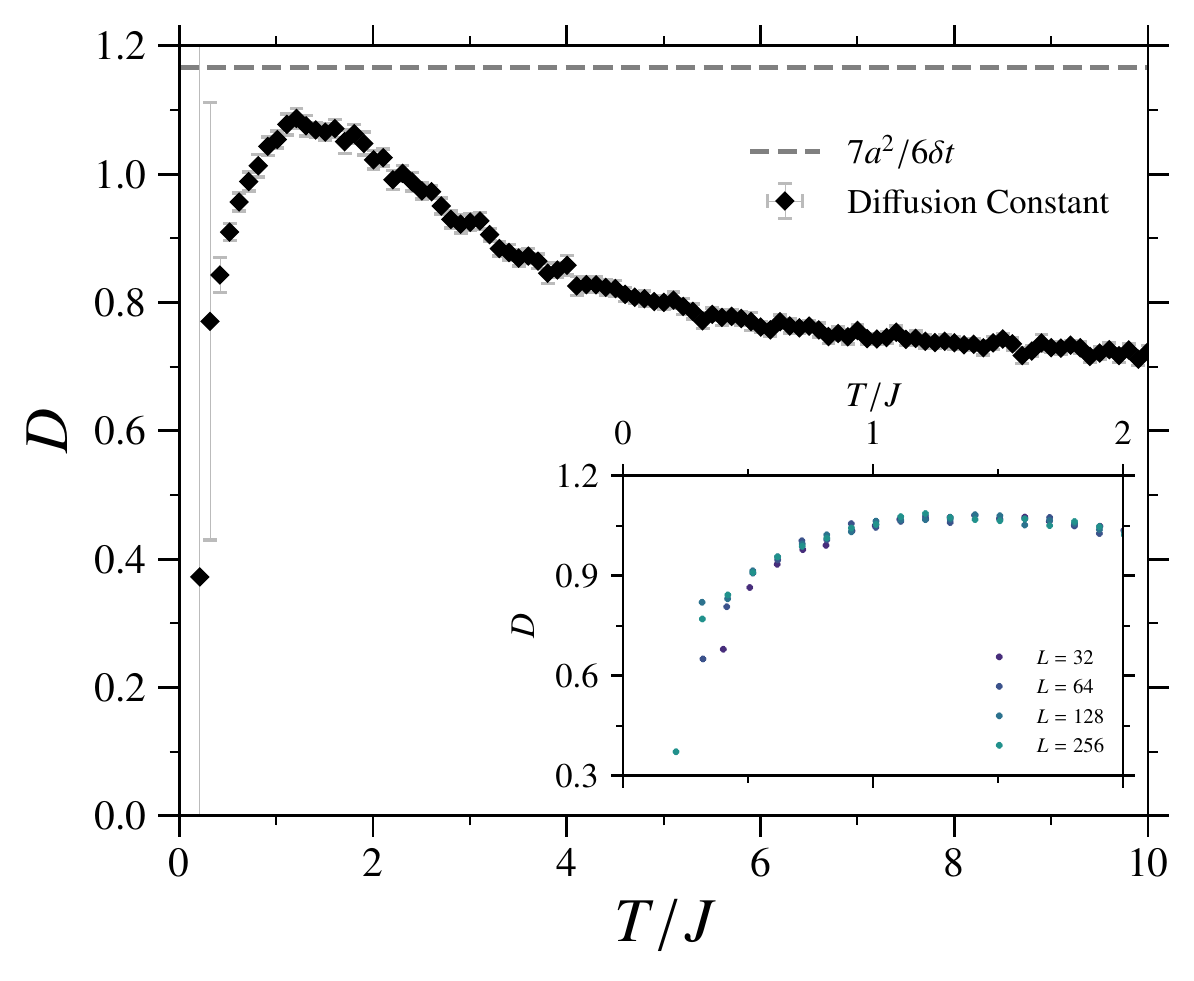}
    \caption{Diffusion constant, $D \equiv \kappa/C$, where $C$ is the heat capacity, as a function of temperature for a $256\times 256 $ system computed using single-spin-flip with demons via the Green-Kubo formula. An estimate for the diffusion constant, $7a^2/6\delta t$, obtained by a random walk through the lattice with some forbidden direction is also shown.  While somewhat constant for moderate to high temperatures, for $T \ll J$ the diffusion constant appears to be approaching zero.
    Natural units with $a$, $J$, $\delta t$ and $k_{\rm B}$ being unity have been used. Note that the error bars become large as $T\rightarrow 0$, this is due to the fact that two small quantities are being divided, as both $\kappa$ and $C$ go to $0$ at low $T$. 
    The temperature dependence of $D$, including the low temperature downturn for $T\lesssim J$, is insensitive to system size $L$, as shown in the in the inset. }
    \label{fig:diffusion-constant}
\end{figure}

The monopole diffusion constant can thus be obtained by re-analyzing the results for $\kappa$ presented in Sec.~\ref{sec:measuring}. For definiteness, we consider single-spin-flip dynamics with demons and look at $\kappa$ obtained for our largest $256\times256$ system from the Green-Kubo formula. We show $D \equiv \kappa/C$ as a function of temperature in Fig.~\ref{fig:diffusion-constant}. From the argument presented above, we would expect that, if the monopoles are diffusive, $D \approx {\rm const.}$ at low temperatures where the density of monopoles is low. We see that $D$ is relatively constant, of $O(1)$, at least for $T \gtrsim J$.

The value of $D$ can be roughly estimated from a simple random walk on the lattice of tetrahedra. Define the diffusion constant for such a random walker with position $\vec{R}(t)$ at time $t$ as the limit
\begin{equation}
\avg{|\vec{R}(t)-\vec{R}(0)|^2} \sim 4 D t,
\end{equation}
as $t \rightarrow \infty$. Elementary arguments (see App.~\ref{app:diffusion}), taking into account distance between the centers of the tetrahedra ($a_M=\sqrt{2}a$) and the average number of hops per time step ($\delta t_M = \delta t/3$) yield $D \approx 3 a^2/(2\delta t)$. Including that one monopole hopping direction is forbidden at each step reduces this by a factor of $7/9$ giving an improved estimate of $D \approx 7a^2/(6\delta t)$ or, in natural units, $D \approx 7/6 = 1.1\bar{6}$. In Fig.~\ref{fig:diffusion-constant} while $D$ is starting to approach this value, at the lowest temperatures $D$ begins to drop precipitously towards zero as $T \rightarrow 0$. 

Furthermore, at low temperature the error bars become large. This is due to the fact that $\kappa$ and $C$ both go to zero exponentially fast, and thus their relative errors are enhanced as $T\rightarrow 0$. This then amplifies the error of $\kappa/C$, leading to the large error bars seen in Fig.~\ref{fig:diffusion-constant}~\footnote{The error in $D$ is calculated using standard error propagation~\cite{taylor1997introduction}, with $\delta D=D\sqrt{(\delta\kappa/\kappa)^2+(\delta C/C)^2}$. This expression could potentially under- or overestimate the errors if the errors in $\kappa$ and $C$ were correlated. Given that the error bars at higher temperatures appear to be small, relative to the noise seen from temperature to temperature we expect the errors have been underestimated. 
}. 

Since this is the regime where we expect the monopole picture to work \emph{best}, we need to examine carefully which of our assumptions has failed. We thus also have computed the monopole diffusion constant $D$ directly, using a method that allows us to approach $T=0$. To this end, consider a system with a single pair of monopole excitations that are allowed to hop, but not annihilate. The mean-squared displacement for one of these monopoles can be written
\begin{equation}
    \label{eq:diffusion-limit}
    D \equiv \lim_{t\rightarrow \infty}\left\{ \frac{\avg{|\vec{R}(t)-\vec{R}(0)|^2}}{4t}
\right\},
\end{equation}
where $\vec{R}(t)$ is the position of the monopole excitation.

Allowing the monopoles to hop, but not annihilate, is identical to the micro-canonical single spin-flip simulations discussed in Sec.~\ref{sec:dynamics:micro-canonical-ssf}~\footnote{The demon method would give the same result, when $T \rightarrow 0$, as the average demon energy is exponentially small.}. More precisely, the system is first initialized in a random ice state before a single spin is flipped at random to create a monopole defect pair. The position of these monopoles is then tracked to create a time series of their position which can be used to compute the mean-square displacement. As is done for auto-correlation functions, the initial time is also averaged over, with
\begin{equation} \label{eq:monopole-position}
\avg{|\vec{R}(t)-\vec{R}(0)|^2}=
\frac{1}{t-t_0}\sum_{t_0=0}^{t}{|\vec{R}(t_0+t)-\vec{R}(t_0)|^2},
\end{equation}
In principle, one should also average over the initial ice state and position of flipped spin, but in practice we find the system to be sufficiently self-averaging so this is unnecessary. 

Practically, the limit $t \rightarrow \infty$ cannot be taken; for displacements of order the system size, $L$, finite-size effects will appear.  If the monopoles are diffusing we would expect the mean-square displacement to be $\propto t$.  Thus, if $D\sim 1$ then we expect these finite-size effects to present themselves at $t \sim (L/4)^2$. Past this time, we expect the monopoles to sample the whole lattice uniformly, yielding a plateau with
$|\vec{R}(t)-\vec{R}(0)|^2 \sim L^2/6$. 

\begin{figure}[tp]
    \centering
    \includegraphics[width=\columnwidth]{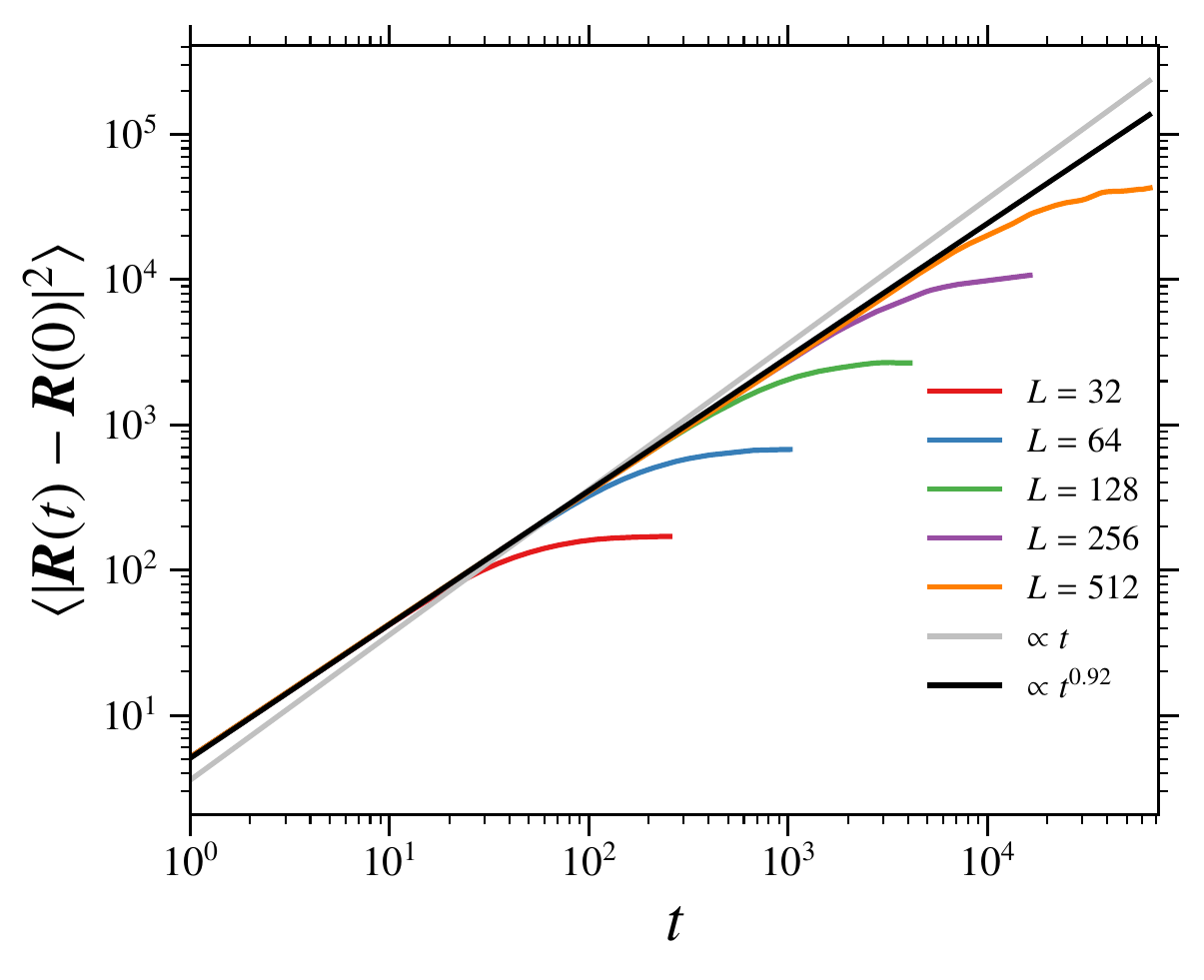}
    \caption{Mean-square displacement for a monopole at $T=0$ as a function of time for several system sizes obtained through Monte Carlo simulation (see Sec.~\ref{sec:diffusion} for details). This deviates from the expected linear time dependence $\propto t$, with slope being $4D$ where $D$ is the monopole diffusion constant. Fitting a power-law $\propto t^\alpha$, the monopole motion is subdiffusive, with a mean-square displacement $\propto t^{0.92}$ being consistent with the data.
    }
    \label{fig:monopole-diffusion}
\end{figure}

Results for the mean-square displacement are shown in Fig.~\ref{fig:monopole-diffusion} for several different system sizes. Although we expect the monopoles to exhibit diffusive behaviour, it is clear that this is not the case. While the mean-square displacement is growing with time, the \emph{power-law} of that growth is lower than the expected value of one. We thus have that monopole motion as $T\rightarrow 0$ is \emph{subdiffusive} with $\avg{|\vec{R}(t)-\vec{R}(0)|^2}\propto t^{\alpha}$ where $\alpha<1$. By fitting a power-law  for times $t\ll(L/4)^2$, one can estimate $\alpha \approx 0.92$. Furthermore, using the definition of $D$ given in Eq.~(\ref{eq:diffusion-limit}) the diffusion constant in fact \emph{vanishes} as $T\rightarrow 0$.

\section{Discussion}
\label{sec:discussion}
 In this section, we address the origin of the subdiffusive monopole motion and comment on whether monopole transport is likely to explain the (magnetic) thermal conductivity in spin ice materials.

\subsection{Origin of subdiffusive behaviour}
\label{sec:discussion:subdiffusive}
\begin{figure}[tp]
    \centering
    \includegraphics[width=\columnwidth]{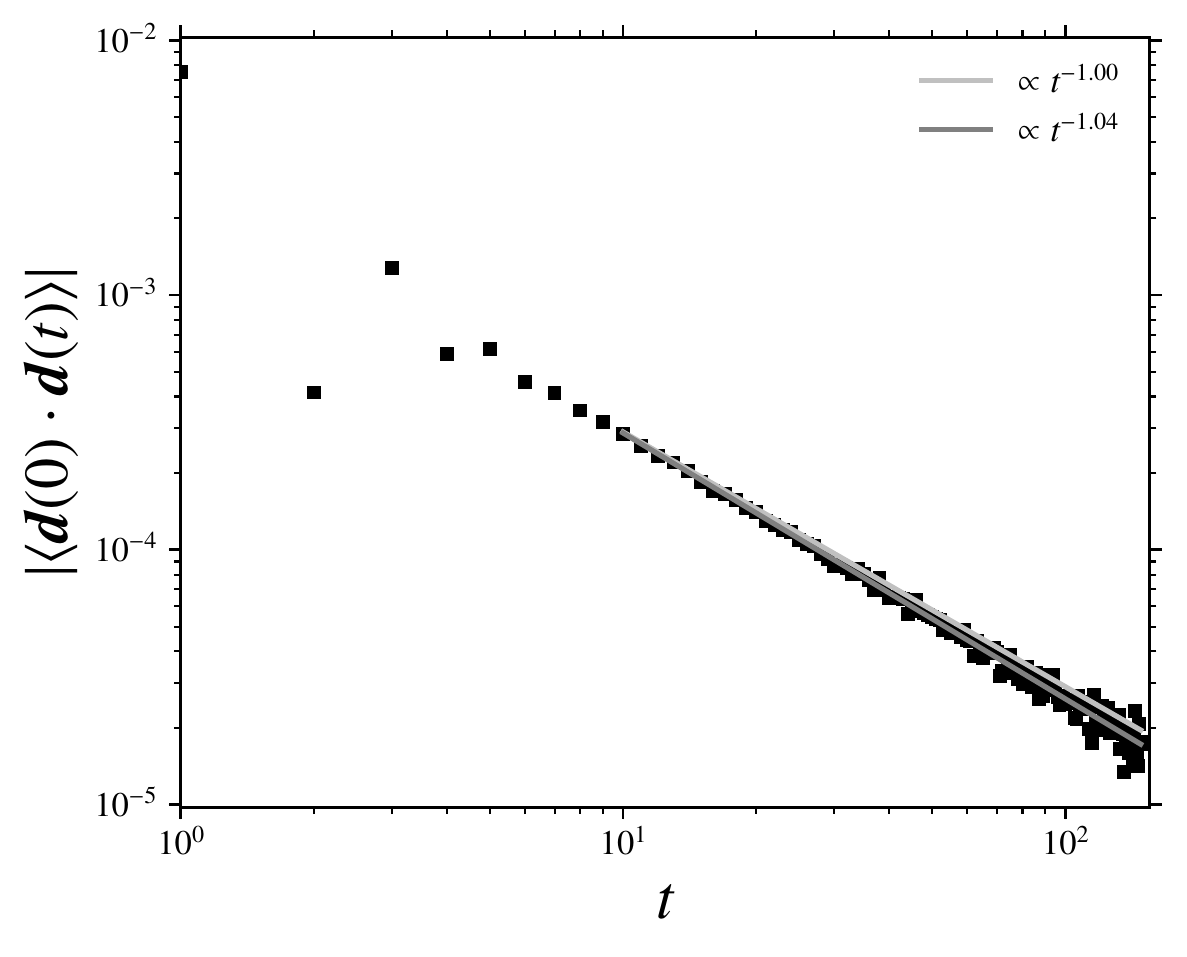}
    \caption{Step auto-correlation function, $\braket{\vec{d}(0)\cdot\vec{d}(t)}$, for a monopole as a function of time, as defined in Eq.~(\ref{eq:step-auto}), calculated using the methods described in Sec.~\ref{sec:diffusion} for a $64\times 64$ system. At long-times its decay is algebraic; fitting a power law yields a scaling of the form $\sim t^{-1.04}$. This is consistent with the subdiffusive monopole motion seen in Fig.~\ref{fig:monopole-diffusion}.
    }
    \label{fig:step-correlator}
\end{figure}

The absence of diffusion, i.e. $\avg{|\vec{R}(t)-\vec{R}(0)|^2} \propto t$ at long times must be traceable to the influence of the spin ice background on allowed monopole hopping paths. The possibility of subdiffusion in square ice was discussed in \citet{nisoli2021color} based on an ansatz for a ``memory function'' associated with monopole hops that scaled algebraically with time. Using this ansatz, they estimated scaling of the mean-squared displacement $\propto t^{\beta-1}$ where $\beta \sim 1.89$, briefly reporting numerical results consistent with this value.

To confirm that the subdiffusive behaviour is originating from memory effects in the monopole hopping process, we examine the correlations between the monopole step directions in detail. In particular, we look at the step auto-correlation function
\begin{equation}
\label{eq:step-auto}
\braket{\vec{d}(0)\cdot\vec{d}(t)} \equiv \frac{1}{t-t_0} \sum_{t_0=0}^{t-1}
\vec{d}(t_0)\cdot\vec{d}(t_0+t),
\end{equation}
where $\vec{d}(t) \equiv \vec{R}(t+1)-\vec{R}(t)$ is the step taken by the monopole at time $t$ and initial times $t_0$ are averaged over. If this function decays sufficiently quickly then diffusive behaviour is expected; for an ideal random walk we expect $\braket{\vec{d}(0)\cdot\vec{d}(t)} \propto \delta_{t,0}$. The result for the step auto-correlation function is shown in Fig.~\ref{fig:step-correlator}, computed using the same methods outlined in Sec.~\ref{sec:diffusion}. This quantity indeed scales algebraically at long time, decaying as $\sim 1/t^{1.04}$, with an exponent slightly larger than one~\footnote{From the (linear) fitting procedure, one can roughly estimate error bars of $-1.04 \pm 0.02$ for this exponent.}. Following the notation of \citet{nisoli2021color}, this corresponds to an exponent of $\beta/2-2$, yielding $\beta \sim 1.92$. This is consistent with their estimate for $\beta$, as well as our estimate from the mean-squared displacement $\beta = 1+\alpha \sim 1.92$ (from Sec.~\ref{sec:diffusion}).

We thus conclude that the subdiffusive behaviour of the monopole motion, and thus the vanishing of $\kappa/C$ as $T \rightarrow 0$, originates from the algebraic correlations of the monopole hoppings, as proposed in Ref.~[\onlinecite{nisoli2021color}]. A more complete analytic and physical understanding of this exponent however is still lacking. Note that some scaling exponents related to monomers in fully packed loop models (equivalent to the six-vertex model) are known~\cite{batchelor1996critical,kondev1997} and may be connected to the value observed in the scaling of the mean-squared displacement.

\subsection{Relevance for Experiments on Spin Ice Materials}
\label{sec:discussion:experiments}
Thermal transport measurements on the spin ice materials Dy$_2$Ti$_2$O$_7$ and Ho$_2$Ti$_2$O$_7$ have revealed a magnetic contribution to the thermal conductivity that appears near the energy scale of the magnetic interactions~\cite{klemke2011thermal,kolland2012thermal,kolland2013anisotropic,toews2013thermal,scharffe2015heat,toews2018}. This magnetic thermal conductivity has been interpreted (at least at low temperatures) by some as being due to the transport of heat by magnetic monopoles~\cite{klemke2011thermal,kolland2012thermal,kolland2013anisotropic,scharffe2015heat}. 

While results for thermal transport in two-dimensional, nearest-neighbour square ice cannot be directly applied to three-dimensional dipolar spin ice, we expect the broad qualitative features to carry over, namely a high temperature $1/T^2$ tail, a peak at $T \sim O(J)$ and a relationship to the monopole diffusion constant as $T \rightarrow 0$.  More quantitatively, the results of Ref.~\cite{samarakoon2021anomalous} show that in three-dimensional spin ice the monopole dynamics are anomalous, that is they do not follow the frequency scaling expected for a diffusive motion. While the precise exponent is close to that expected for an ideal random walk for nearest neighbour spin ice, it becomes more and more anomalous (subdiffusive) as dipolar and exchange corrections are included. We thus expect that, as in the two-dimensional square ice, the thermal diffusivity  (i.e. the monopole diffusion constant) in three-dimensional spin ice vanishes as temperature approaches zero.

Returning to comparison with experiment, first, consider the \emph{magnitude} of the expected magnetic contribution. Since the monopoles are not ballistic, we expect that the sample size is irrelevant and thus an estimate for the bulk thermal conductivity can be made. From Fig.~\ref{fig:baths-w-gradient}, in natural units the maximum value of the thermal conductivity is $\kappa^{*} \sim 0.3$. Restoring time and length scales one finds
$$
\kappa^{*} \sim 0.3 \cdot \frac{a^2 k_{\rm B}}{\delta t V_1},
$$
where $\delta t$ is the spin-flip time, $J$ the exchange constant, $V_1$ volume per spin and $a$ is the nearest neighbour-distance. The length scales are taken as appropriate for spin ice materials: $a \sim 3.5 \AA$, $V_1 \sim 62.5 \AA^3$, assuming a conventional cubic unit cell is $\sim 10\AA$ in size~\cite{gardner2010magnetic}. For the spin-flip time, $\delta t$, first consider the time-scale extracted from magnetic relaxation measurements~\cite{snyder2004low,yaraskavitch2012spin}: $\delta t \sim 1\ {\rm ms}$. This yields the thermal conductivity
$$
\kappa^{*} \sim 10^{-11} \ \frac{\rm W}{\rm K\ m}
$$
This is \emph{far} too small to account for the magnetic thermal conductivity in spin ice materials. For example for Dy$_2$Ti$_2$O$_7$, Ref.~[\onlinecite{kolland2012thermal}] reports a maximum thermal conductivity of $\kappa^* \sim 10^{-1} \cdot {\rm W}/(\rm K\ m)$ -- ten orders of magnitude larger. The drastic difference in scale arises from \emph{slowness} of the magnetic dynamics in spin ice. For comparison, when converted to an energy, the spin-flip time $\delta t$ corresponds to $\hbar/\delta t \sim 10^{-10} \meV$, ten orders of magnitude smaller than typical energy scales in rare-earth magnets.

While the magnitude is wildly incorrect, the temperature dependence of the thermal conductivity is similar to the experimental data~\cite{kolland2012thermal}: there is a high temperature tail, a maximum near $T \sim 1\K$ (similar to the location of the maximum in the heat capacity) and $\kappa$ approaches zero as $T$ goes to zero. The thermal diffusivity, $\kappa/C$, is also approximately temperature independent above $\sim 1\K$ with a value of $\sim 0.25 \cdot 10^{-5} \ {\rm m^2/s}$. 

One na\"ive explanation could then be that the dynamics that are operative for energy transport are simply much faster than expected based on magnetic relaxation. To match the overall scale a spin-flip time of $\delta t \sim 10^2\ {\rm fs}$, or equivalently an energy scale $\hbar/\delta t \sim 10 \meV$, would be needed. However, there is no known magnetic energy of this size in these systems~\footnote{While the gap to the first excited crystalline electric field energy scale \emph{is} of $O(10 {\rm meV})$, this should play no role in the dynamics at low temperature.}. In addition to such a large energy scale, one would also need a mechanism to explain why the magnetic relaxation remains slow, while energy transport is fast. For these reasons, we do not consider this explanation viable.

A more plausible explanation could involve the lattice, with the interaction between the (slow) magnetic degrees of freedom and the (fast) phonon excitations contributing to the magnetic thermal conductivity. For example, scattering of phonons from a distribution of magnetic monopoles (quasi-static on lattice time scales and carrying electric dipole moments~\cite{khomskii2012electric}) or the spin ice background itself would likely affect the phonon contribution to the thermal conductivity. However, scattering from point-like defects (e.g. monopoles) or line-like defects (e.g. loops in the spin ice background) are expected to be insignificant relative to the boundary scattering contribution at sufficiently low temperature~\cite{ziman2001electrons}. 

Alternatively, modification of the sound velocity, and thus the boundary scattering contribution, due to spin-phonon interactions is possible but would have to be quite large to account for the observable magnetic conductivity. In Ref.~[\onlinecite{kolland2012thermal}] one has $\kappa_{\rm mag}/\kappa_{\rm tot} \sim 0.25$ at $1\K$. Since we expect $\kappa \propto 1/v^2$ in the boundary scattering limit, to yield a $25\%$ increase a change of $\Delta v/v \sim -0.1$ would be needed over a range of $\sim 1\K$ or so.  Since measurements of the relative change in the sound velocity~\cite{ultrasonic} in Dy$_2$Ti$_2$O$_7$ and Ho$_2$Ti$_2$O$_7$ are of order $|\Delta v/v| \sim 10^{-5}$ (and not always the correct sign), we take this explanation as unlikely.

Finally, note that the two-dimensional square ice model can be more directly related to \emph{artifical} spin ice~\cite{nisoli2013rmp,skjaervo2020advances}: arrays of nanomagnets (ferromagnetic islands) engineered to mimic the physics of spin ice. However, while the motion of monopole defects has been studied in such systems~\cite{pollard2012,kapaklis2014thermal,farhan2019emergent,arava2020}, it is not clear that thermal conductivity is a useful or practical quantity to measure. Indeed, the information encoded in ``macroscopic'' quantities such as $\kappa$ can likely be accessed more directly given the detailed magnetic configuration which may be imaged directly in space and as a function of time.

\section{Conclusion}
\label{sec:conclusion}
In summary, we have explored thermal transport in square ice using energy-conserving kinetic Monte Carlo methods. We computed thermal conductivity using two different kinds of stochastic dynamics, showing that the qualitative features are independent of the dynamics used. Furthermore, at low temperatures, we found that energy is transported by magnetic monopoles, with the thermal diffusivity vanishing as we approach $T=0$ due to subdiffusive motion of the monopole excitations. We explored the origins of this subdiffusive behaviour and related them to algebraic correlations between the monopole step directions. Finally, we concluded that monopole transport due to single-spin-flip dynamics cannot account for the magnetic contribution to the thermal conductivity in spin ice materials.

We close with a few of the many important open questions regarding thermal transport in spin ice, and in thermal transport in frustrated magnets more broadly. 

First, in spin ice, understanding the origin of the magnetic thermal conductivity in Dy$_2$Ti$_2$O$_7$ and Ho$_2$Ti$_2$O$_7$ needs further study~\cite{klemke2011thermal,kolland2012thermal,kolland2013anisotropic,toews2013thermal,scharffe2015heat,toews2018}. In particular, the role of the lattice and spin-lattice coupling in producing an apparent magnetic contribution to the thermal conductivity remains somewhat elusive. A detailed accounting of the effects of phonon scattering (resonant and non-resonant) from the magnetic degrees of freedom, and its interplay with the evolution of the spin ice physics, is likely necessary to determine the viability of such explanations.

Second, from a broader perspective, while we have focused here on square ice, this methodology can be applied to other frustrated models that exhibit different kinds of ground state manifolds or fractionalized excitations. Examples include models of classical $Z_2$ spin liquids~\cite{PhysRevLett.118.047201}, models exhibiting fragmentation~\cite{lhotel2020fragmentation} or those realizing higher-rank gauge theories~\cite{pretko2020fracton}, to name only a few. Understanding the mechanisms of energy transport in these systems, and the role played by their fractionalized excitations, offers exciting opportunities for future study. 

Finally, another natural line of questioning concerns the effect of magnetic fields, and consequently the appearance of the thermal Hall conductivity, $\kappa_{xy}$. Puzzling measurements of the thermal Hall effect in the Kitaev magnet RuCl$_3$~\cite{rucl1,rucl2,kasahara2018majorana} and in pyrochlores such as Tb$_2$Ti$_2$O$_7$~\cite{hirschberger2015large} and Yb$_2$Ti$_2$O$_7$~\cite{hirschberger2019enhanced} have sparked renewed interest in $\kappa_{xy}$ and presented new challenges to theory. More fundamentally, ambiguities in the definition of the bulk thermal Hall conductivity~\cite{cooper1997,kapustin2020} arise even in this classical context (see App.~\ref{app:green-kubo}), and thus developing methods for accessing $\kappa_{xy}$ is an important goal. 

It is clear that there is significant work left to be done in understanding thermal transport in frustrated magnets. This includes both spin ice models specifically, as well as in the large and varied family of models and materials that exhibit high frustration. We hope that the results and methods presented here will further motivate theoretical and experimental studies of these systems.

\begin{acknowledgements}
We thank Michel Gingras and Ludovic Jaubert  for helpful comments and discussions. We acknowledge the support of the Natural Sciences and Engineering Research Council of Canada (NSERC), [funding reference number RGPIN-2020-04970].
\end{acknowledgements}

\appendix

\section{Energy Current}
\label{app:energy-current}
\newcommand{\eq}{{\rm eq}}
\renewcommand{\s}{\vec{\sigma}}
In this appendix, we derive the energy current given in Eq.~(\ref{eq:local-energy-current}) of the main text. Consider a general (nearest-neighbour) Ising model with a ``demon'' bath, as in Sec.~\ref{sec:dynamics:ssf-demons}
$$
E[\s,\vec{D}] = J \sum_{\braket{ij}} \sigma_i \sigma_j + 
\sum_i D_i,
$$
where $\sigma_i = \pm 1$ and $D_i = 0,4J,8J,\dots$. For a given configuration of spins
at time $t$, the spins and demons are updated via a local flip. A spin $k$ is picked at random from the lattice, the spin-flip cost $\Delta E_k$ is computed [Eq.~(\ref{eq:spin-flip-energy})] and the update is performed as described in Sec.~\ref{sec:dynamics:ssf-demons}. This can be written as
\begin{align*}
    \sigma_k'&= \begin{cases}
    +\sigma_k, & D_k < \Delta E_k, \\
    -\sigma_k, & D_k \geq \Delta E_k,
    \end{cases},\\
    D_k' &= \begin{cases}
    D_k, & D_k < \Delta E_k, \\
    D_k-\Delta E_k, &  D_k \geq \Delta E_k,
    \end{cases},
\end{align*}
where $\vec{\sigma}'$ and $\vec{D}'$ are the updated spin and bath variables and
$\sigma_i$ and $D_i$ are unchanged for $i\neq k$. By construction, the total energy is conserved under these updates, with
$$
E[\s',\vec{D}'] = 
E[\s,\vec{D}].
$$
This can be written a bit more compactly in terms of a Heaviside function 
\begin{subequations}
\label{eq:app:update}
\begin{align}
    \sigma_i' &= \sigma_i \left[1-2\delta_{ik}
    \Theta_{D_i \geq \Delta E_k}\right],\\
    D_i' &= D_i - \delta_{ik}
    \Theta_{D_i \geq \Delta E_k} \Delta E_k,
\end{align}
\end{subequations}
where $\Theta_C=1$ if $C$ is true and $\Theta_C = 0$ otherwise.

Consider now a local energy density and how that defines energy flow under these dynamics. Write the energy density
$$
\epsilon_i \equiv D_i + \frac{1}{2} J \sigma_i \sum_{\braket{i \rightarrow j}} \sigma_j,
$$
so that $E = \sum_i \epsilon_i$. Note the factor of $1/2$ to account for double-counting the bonds; i.e. $\sum_{\braket{ij}} \equiv \frac{1}{2} \sum_i \sum_{\braket{i\rightarrow j}}$. 
One can then obtain the change after one step using Eq.~(\ref{eq:app:update})
\begin{align*}
\epsilon_i' &\equiv D_i' + \frac{1}{2} J \sigma_i' \sum_{\braket{i 
\rightarrow j}} \sigma_j',   \\ 
 &= \epsilon_i 
    -      \Theta_{D_k \geq \Delta E_k}\left\{ \delta_{ik} \Delta E_k+
    J\sum_{\braket{i j}} 
    \sigma_i \sigma_j
    \left[
    \delta_{ik}+
    \delta_{jk}
    \right]
    \right\},
\end{align*}
where $\epsilon_i'$ is the new energy density and that $i \neq j$ has been exploited to remove a few terms.
This can be written more suggestively if $\Delta E_k$ is expanded out
\begin{align}
\epsilon_i' -\epsilon_i
 &= -\sum_{\braket{i \rightarrow j}} \left[
    J\Theta_{D_k \geq \Delta E_k}
    \sigma_i \sigma_j
    \left(\delta_{jk}-\delta_{ik}\right)
    \right]. \nonumber
\end{align}
Using the continuity equation to equate this to $-\sum_{\avg{i \rightarrow j}} j^{(k)}_{ij}$ the energy current
for this update [as in Eq.~(\ref{eq:local-energy-current})] can be identified
\begin{equation}
 j_{ij}^{(k)} \equiv J\Theta_{D_k \geq \Delta E_k}
    \sigma_i \sigma_j
    \left(\delta_{jk}-\delta_{ik}\right),
\end{equation}
where this quantity is only non-zero for nearest neighbours. To obtain $j_{ij}(t)$, these contributions are summed over a full sweep of $N$ randomly chosen sites $k$ (with the spin configuration updated after each accepted flip).

\section{Green-Kubo Formula}
\label{app:green-kubo}
In this appendix, an overview of the derivation of the Kubo formula for the thermal conductivity in a (kinetic) Ising model is provided. We follow the treatment presented in \citet{saito1999transport}, save for the evaluation of the $\avg{\vec{J}}_{0}$ term, for which we provide a different, somewhat more general argument that yields the symmetric parts of the thermal conductivity matrix.

Consider a system with states defined by
$\s = (\sigma_1,\sigma_2,\dots,\sigma_N)$. We assume that 
we have a state $\s(t)$ at each time $t$. The evolution of the
state from time $t$ to $t+1$ takes the form
$$
\s(t+1) \equiv \Omega(\s(t)),
$$
where $\Omega$ encapsulates a (memoryless)  updating process~\footnote{Strictly we derive the Green-Kubo formula for a deterministic process that is time independent. Random picking of the spin violates strictly violates this, as the process defined by $\Omega$ actually depends on time $t$. However, since the updates are time-independent ``on average'' we do not expect any issues in applying this formalism.}. We assume that an energy $E(\s)$ that is preserved by $\Omega$, with $E(\Omega(\s)) = E(\s)$, can be defined. The time evolution of any observable $O(\s) \equiv O_0(\s)$ starting from state $\s$ can then be defined via
$$
O_{t+1}(\s) \equiv O_t(\Omega(\s)).
$$

We assume that this dynamics will, in the long time limit, sample from the equilibrium distribution
\begin{equation}
    \label{eq:eqdist}
    P_{\eq}(\s) = \frac{e^{-\beta E(\s)}}{\sum_{\s'} e^{-\beta E(\s')}} ,
\end{equation}
for some inverse temperature $\beta \equiv 1/(k_{\rm B} T)$ starting from some (essentially) arbitrary initial state. By construction, this distribution is assumed to be invariant under the chosen dynamics, with $P_{\eq}(\Omega(\s)) = P_{\eq}(\s)$ for any possible inverse temperature $\beta$ since the energy, $E(\s)$, is invariant.

To probe thermal transport, assume the initial state $\s(0)$ is drawn from a distribution with a thermal gradient
\begin{equation}
P_0(\s) \equiv \frac{e^{-\sum_i \beta_i \epsilon_i(\s)}}{\sum_{\s'} e^{-\sum_i \beta_i \epsilon_i(\s)}},
\end{equation}
where $\beta_i$ is a distribution of local inverse temperatures that encodes the gradient and $\epsilon_i$ is the local energy density, which decomposes as $E = \sum_i \epsilon_i$. Explicitly, write $\beta_i = 1/(k_{\rm B} T_i)$ with
$$
T_i \equiv T + \vec{r}_i \cdot (\grad T) ,
$$
where $T$ is the mean temperature and $\grad T$ is constant.
Denote the statistical averages with respect to $P_0$ as
$$
\avg{O} \equiv \sum_{\s} O(\s) P_0(\s).
$$
Since the dynamics preserves $E(\s)$, this distribution of states will \emph{not} be preserved as time evolves. Define the time evolved distribution as $P_t$ with
$$
P_{t+1}(\s) = \sum_{\s'} \delta_{\s,\Omega(\s')}
P_t(\s').
$$
We are mainly interested in the statistical averages of observables as a function of time, denoted as
$$
\avg{O_t} \equiv \sum_{\s} P_0(\s) O_t(\s),
$$
for the average of $O$ over an initial ensemble of states distributed as $P_0$ that have been evolved for a time $t$.
Equivalently, the time-dependence in the observable $O_t$ can be traded for the time-dependence in the probability distribution $P_t$ with
$$
 \avg{O_t} = \sum_{\s} P_t(\s) O(\s).
$$
This can also be done partially, e.g. with
$
 \avg{O_t} =  \sum_{\s} P_m(\s) O_{t-m}(\s)     
$ for any $m \geq t$.

Consider now the expectation of the energy current in the presence of this temperature gradient. The current is related to the energy density via the a discrete continuity equation [Eq.~(\ref{eq:continuity})]
\begin{equation}
    \label{eq:contin}
    \frac{\del \epsilon_i(\vec{\sigma})}{\del t} \equiv \frac{\epsilon_i(\Omega(\s)) - \epsilon_i(\s)}{\delta t} = -\sum_j j_{ij}(\s),
\end{equation}
where $j_{ij} = -j_{ji}$ defines a local energy current flowing from site $i$ to $j$ and $\delta t$ the duration of the discrete time step. The total current can be defined as [Eq.~(\ref{eq:total-current})]
\begin{equation}
\label{eq:curr}
\vec{J} \equiv \frac{1}{2} \sum_{ij} (\vec{r}_j -\vec{r}_i) j_{ij}.
\end{equation}

Our goal is to compute the thermal conductivity, $\kappa_{\mu\nu}$, as given in Eq.~(\ref{kappa-general}) which is defined as the response of the energy current density to a small thermal gradient. We will be more precise here, paying attention to the order of the long time and large system size limits
\begin{equation}
\label{eq:kappa}
\lim_{t \rightarrow \infty} \lim_{V\rightarrow \infty} \left[\frac{1}{V}  \avg{J^\mu_t}\right] \equiv -\sum_{\nu} \kappa_{\mu\nu} (\grad T)_{\nu} + \cdots.
\end{equation}
Strictly, the limit of infinite volume $V \rightarrow \infty$ needs to be taken first (keeping $N/V$ fixed), and only then $t \rightarrow \infty$ to prevent the system from reaching equilibrium. 

There are two time scales at play: first, there is time for the initial thermal gradient to induce a steady, flowing energy current that allows determination of the thermal conductivity -- call this $\tau_{\textrm{flow}}$. The second scale is the thermalization time for the whole system; eventually the system will clear the thermal gradient and adopt some uniform temperature -- call this $\tau_{\textrm{therm}}$.  For a large system, the thermalization time $\tau_{\textrm{therm}}$ should scale with the system size, schematically like $\tau_{\textrm{therm}} \sim N^\alpha$, and thus if $N\rightarrow \infty$ is taken first then $\tau_{\textrm{therm}} \gg \tau_{\textrm{flow}}$. Pragmatically, for a finite system the statement of $\lim_{t\rightarrow \infty}$ thus means that $\tau_{\textrm{flow}} \ll t \ll \tau_{\textrm{therm}}$.

\subsection{Perturbation theory for $\avg{J^\mu_t}$}
Following the strategy  of \citet{saito1999transport}, look at the evolution of the total current $J^\mu$ along some direction by breaking up its evolution into a sum of discrete changes
\begin{align*}
\avg{J^\mu_t} &= \avg{J^\mu_0} + \sum_{\tau=0}^{t-1} \avg{J^\mu_{\tau+1}-J^\mu_{\tau}}, \\
&= \avg{J^\mu_0} + \sum_{\tau=0}^{t-1} \sum_{\s} 
\left[J^\mu_{\tau+1}(\s)-J^\mu_{\tau}(\s)\right]P_0(\s), \\
&= \avg{J^\mu_0} + \sum_{\tau=0}^{t-1} \sum_{\s} J^\mu_\tau(\s) \left[P_1(\s)-P_0(\s)\right].
\end{align*}
Consider writing $P_1(\s)$ explicitly via the continuity equation [Eq.~(\ref{eq:contin})]. The energy in the presence of the thermal gradient can be re-written as
$$
\sum_i \beta_i \epsilon_i(\s) = 
\sum_i \beta_i \epsilon_i(\Omega(\s))
+\delta t\sum_i \beta_i  \sum_{j} j_{ij}(\s) .
$$
Putting this into the definition of $P_1$ yields
$$
P_1(\vec{\sigma}) = \sum_{\s'} \delta_{\s,\Omega(\s')}
P_0(\s') = P_0(\s)
 \sum_{\s'} \delta_{\s,\Omega(\s')}
e^{-\delta t \sum_i \beta_i  \sum_{j} j_{ij}(\s') },
$$
which, using that $j_{ij} = -j_{ji}$, simplifies to
$$
\sum_{ij} \beta_i j_{ij}(\s') = 
\frac{1}{2}\sum_{ij} \left(\beta_i-\beta_j\right) j_{ij}(\s'),
$$
depending on the difference of the inverse temperatures. Assuming a small temperature gradient
$$
 \beta_i-\beta_j  \approx
  -\frac{1}{k_{\rm B} T^2} \left(\vec{r}_i-\vec{r}_j\right) \cdot \grad T + \cdots,
$$
gives an expression involving the total current $\vec{J}$
$$
\frac{1}{2}\sum_{ij} \left(\beta_i-\beta_j\right) j_{ij}(\s') = 
+\frac{1}{k_{\rm B} T^2}(\grad T) \cdot \vec{J}(\s').
$$
The quantity $P_1$ can then be expanded to leading order in $\grad T$ to obtain
\begin{equation*}
P_1(\vec{\sigma})-P_0(\vec{\sigma}) \approx 
-P_0(\s)
\delta t \beta^2 (k_{\rm B} \grad T) \cdot
\left[
 \sum_{\s'} \delta_{\s,\Omega(\s')} \vec{J}(\s')
\right],
\end{equation*}
where we have used that $\sum_{\s'} \delta_{\s,\Omega(\s')}=1$. Further, in this expression $P_{0}$ can be replaced with $P_{\rm eq}$, as corrections are higher order in $\grad T$, yielding
$$
\avg{J^\mu_t} = 
\avg{J^\mu_0} -\delta t \beta^2 (k_{\rm B} \grad T) \cdot \sum_{\tau=0}^{t-1} \sum_{\s} J^\mu_\tau(\s) 
P_{\eq}(\s)
\left[
 \sum_{\s'} \delta_{\s,\Omega(\s')}\vec{J}(\s)
\right].
$$
Since the equilibrium distribution is invariant under the chosen dynamics, this can be simplified
\begin{align*}
   \sum_{\s} J^\mu_\tau(\s) 
P_{\eq}(\s)\delta_{\s,\Omega(\s')} &= 
P_{\eq}(\s')\sum_{\s} J^\mu_\tau(\s) 
\delta_{\s,\Omega(\s')}, \\
&= P_{\eq}(\s')J^\mu_{\tau+1}(\s').
\end{align*}
This then yields
\begin{align*}
\avg{J^\mu_t} &= 
\avg{J^\mu_0} - \delta t \beta^2 (k_{\rm B} \grad T) \cdot \sum_{\tau=0}^{t-1}  
 \sum_{\s}  
 P_{\eq}(\s)J^\mu_{\tau+1}(\s) \vec{J}(\s),    \\
 &= 
\avg{J^\mu_0} - \delta t \beta^2 \sum_{\nu} (k_{\rm B} \nabla_\nu T) \sum_{\tau=1}^{t}  
 \avg{J^\mu_{\tau} J^\nu_0}_{\eq},
\end{align*}
where in the last term the sum has been shifted by one.

\subsection{Evaluation of $\avg{J^\mu_0}$}
The first term, $\avg{\vec{J}_0}$, which involves the distribution with local temperatures, still needs to be evaluated. While we cannot evaluate this in general, we can extract the symmetric part of the thermal conductivity matrix. 
First, consider the expansion in exponential in powers of $\grad T$
\begin{align*}
    \sum_i \beta_i \epsilon_i(\s) &\approx
\beta \sum_i \epsilon_i(\s) -  \beta^2 (k_{\rm B} \grad T) \cdot \sum_i \vec{r}_i \epsilon_i(\s) +\cdots,\\
&\equiv \beta E(\s) - \Phi(\s).
\end{align*}
A series expansion in $\Phi \propto \grad T$ in the numerator and denominator yields
$$
\avg{J^\mu} = \frac{\sum_{\s} J^\mu(\s) e^{-\beta E(\s) + \Phi(\s)}}{\sum_{\s} e^{-\beta E(\s) + \Phi(\s)}} \approx \avg{J^\mu}_{\eq} + \left(\avg{J^\mu \Phi}_{\eq}-\avg{J^\mu}_{\eq}\avg{\Phi}_{\eq}\right).
$$
Since $\avg{J^\mu}_{\eq} =0$, one is left with
$$
\avg{J^\mu} =  \beta^2 (k_{\rm B} \grad T) \cdot \avg{J^\mu \vec{P} }_{\eq},
$$
where the energy polarization is defined as $\vec{P} \equiv \sum_i \vec{r}_i \epsilon_i$ (not to be confused with the probability distribution).

\subsection{Symmetric Part of Thermal Conductivity}
The expectation value $\avg{J^\mu}$ is proportional to $\grad T$ and thus it contributes to the thermal conductivity. Using the definition of $\kappa$ in Eq.~(\ref{eq:kappa}) we thus have
\begin{equation}
\kappa_{\mu\nu} = \frac{1}{k_{\rm B} V T^2}\left\{
- \avg{J^\mu P^\nu }_{\eq}+
\delta t \sum_{\tau=1}^{\infty}\avg{J^\mu_{\tau} J^\nu_0}_{\eq} .
\right\}
\end{equation}
For the symmetric part of $\kappa_{\mu\nu}$, defined as $\bar{\kappa}_{\mu\nu} \equiv (\kappa_{\mu\nu}+\kappa_{\nu\mu})/2$, the expectation $\avg{J^\mu P^\nu}$ can be related to the current-current average $\avg{J^\mu J^\nu}$. Explicitly, we prove the identity
\begin{equation}
\label{eq:JJ-identity}
\delta t \avg{J^\mu J^\nu}_{\eq} = -\avg{J^\mu P^\nu}_{\eq}
-\avg{J^\nu P^\mu }_{\eq}.
\end{equation}
\begin{widetext}
To see how this is true, first express $\vec{J}$ using the continuity equation Eq.~(\ref{eq:contin})
$$
\sum_i \vec{r}_i \left(\epsilon_i(\Omega(\s))-\epsilon_i(\s)\right)=
-\delta t \sum_{ij} \vec{r}_i j_{ij}(\s)=
-\frac{1}{2}\delta t\sum_{ij} \left(\vec{r}_i-\vec{r}_j\right) j_{ij}(\s) = +\delta t\vec{J}(\s).
$$
This then yields 
\begin{align*}
\delta t^2\avg{J^\mu J^\nu}_{\eq}    &=
\sum_{\s} \sum_{ij} r^\mu_i (\epsilon_i(\Omega(\s))-\epsilon_i(\s))
 r^\nu_j (\epsilon_j(\Omega(\s))-\epsilon_j(\s)) P_{\eq}(\s), \\
 &=
 \sum_{\s} \sum_{ij} r^\mu_i r^\nu_j\left[
  \epsilon_i(\Omega(\s))[\epsilon_j(\Omega(\s))-\epsilon_j(\s)]-
  \epsilon_i(\s)[\epsilon_j(\Omega(\s))-\epsilon_j(\s)]
  \right] P_{\eq}(\s).
\end{align*}
Some of these terms can be simplified using the time-invariance of the equilibrium distribution
$$
 \sum_{\s} \sum_{ij} r^\mu_i r^\nu_j
  \epsilon_i(\Omega(\s))\epsilon_j(\Omega(\s))P_{\eq}(\s) =
   \sum_{\s} \sum_{ij} r^\mu_i r^\nu_j
  \epsilon_i(\s)\epsilon_j(\s)P_{\eq}(\s).
$$
Putting this back in allows regrouping of terms as
\begin{align*}
\delta t^2 \avg{J^\mu J^\nu}_{\eq}    
 &=
 -\sum_{\s} \sum_{ij} r^\mu_i r^\nu_j\left[
  \epsilon_j(\s)[\epsilon_i(\Omega(\s))-\epsilon_i(\s)]+
  \epsilon_i(\s)[\epsilon_j(\Omega(\s))-\epsilon_j(\s)]
  \right] P_{\eq}(\s), \\
 &=
 -\delta t\left\{\sum_i \avg{r^\mu_i \epsilon_i J^\nu}_{\eq}  
 +\sum_i \avg{r^\nu_i \epsilon_i J^\mu}_{\eq}\right\},
\end{align*}
so the identity in Eq.~(\ref{eq:JJ-identity}) has been proved.
\end{widetext}
Putting this together, the symmetric part is given by
\begin{align*}
   \bar{\kappa}_{\mu\nu} &= \frac{1}{k_{\rm B} V T^2}\left\{
\frac{1}{2}\delta t \avg{J^\mu_0 J^\nu_0}_{\eq}+
\frac{1}{2}\delta t \sum_{\tau=1}^{\infty}\avg{{J^\mu_{\tau} J^\nu_0+J^\nu_{\tau} J^\mu_0}}_{\eq} 
\right\}.
\end{align*}
The first term can be absorbed into the sum by including a $\tau=0$ correction,
giving the final result~\footnote{This can be simplified somewhat by re-writing $\avg{J^\nu_\tau J^\mu_0}_{\eq}=\avg{J^\nu_0 J^\mu_{-\tau}}_{\eq}$ then assuming a time-reversal symmetry so that $\avg{J^\nu_0 J^\mu_{-\tau}}_{\eq}=\avg{J^\mu_\tau J^\nu_0}_{\eq}$ but this is not true in general (e.g. in any setup where the thermal Hall conductivity is finite).
}
\begin{equation}
\bar{\kappa}_{\mu\nu} = 
 \frac{\delta t}{2k_{\rm B} VT^2}\sum_{\tau=0}^{\infty}  
 \avg{J^\mu_{\tau} J^\nu_0+J^\nu_{\tau} J^\mu_0}_{\eq} \left(1-\frac{1}{2}\delta_{\tau,0}\right).
\end{equation}
The diagonal parts are simpler, with the symmetrization of the current correlator being unnecessary
$$
{\kappa}_{\mu\mu}= \frac{\delta t}{k_{\rm B} VT^2}\sum_{\tau=0}^{\infty}  
 \avg{J^\mu_{\tau} J^\mu_0}_{\eq} \left(1-\frac{1}{2}\delta_{\tau,0}\right).
$$

Note that the \emph{anti}-symmetric part, which encodes the thermal Hall effect, is not accessible through this strategy, since the energy polarization is difficult to define as a bulk quantity. Evaluating the anti-symmetric components requires dealing with the non-transport part of the energy currents, i.e. the energy magnetization (see, e.g., \cite{cooper1997,kapustin2020}).

\section{Estimates of Monopole Diffusion Constant}
\label{app:diffusion}
In this appendix we review some simplest estimates for the monopole diffusion constant. Consider a monopole on a square lattice with lattice constant $a_M$ that starts at the origin at $t=0$. After each interval of time $\delta t_M$ the particle takes a step to one of its nearest neighbours at random, in the directions $+\vhat{x}$, $+\vhat{y}$, $-\vhat{x}$ and $-\vhat{y}$ with equal probability. Write the (random) position at time $t$ as 
$$
\vec{R}(t) = \sum_{n=0}^{N_M} \vec{d}_n,
$$
where $N_M \equiv t/\delta t_M \gg 1$. We assume each of the (random) $\vec{d}_n$ is drawn independently. The expected mean-squared displacement from the origin is given by
$$
\avg{|\vec{R}(t)|^2} = \sum_{n,n'=0}^{N_M} \avg{\vec{d}_n\cdot \vec{d}_{n'}} =
\sum_{n=0}^{N_M} \avg{|\vec{d}_n|^2} = N_M a_M^2 = \frac{a_M^2 t}{\delta t_M}.
$$
Here we have used that independence of the steps implies that $\avg{\vec{d}_n\cdot \vec{d}_{n'}} = \delta_{nn'} 
\avg{|\vec{d}_n|^2}$ and that $|\vec{d}_n|^2=a_M^2$ for all possible steps.

If the monopoles in square ice are considered as diffusing randomly on the dual square lattice formed by the tetrahedra then $a_M = \sqrt{2} a$ where $a$ is the nearest neighbour spacing.  The hopping time $\delta t_M$ needs to be related to the spin-flip time; since each monopole can hop by flipping one of three different spins on its tetrahedron, we expect that $\delta t_M = \delta t/3$ where $\delta t$ is time for complete one attempted flip per spin (one sweep). Using the most na\"ive approximation then yields
$$
\avg{|\vec{R}(t)|^2}  = \frac{6 a^2 t}{\delta t}.
$$
In two dimensions one has $\avg{|\vec{R}(t)|^2} \sim 4Dt$ at long times where $D$ is the monopole diffusion constant. Therefore this would give the estimate $D \sim 3 a^2/(2 \delta t)$, or, in natural units with $a=\delta t=1$, a diffusion constant of $D = 3/2$.

However, in square ice for each hop one of the four possible directions is disallowed. The disallowed direction is also \emph{never} the direction that would cause the particle to backtrack and so the hops are not uncorrelated. This estimate for $D$ can be improved by incorporating some of this physics. Explicitly, assume that this disallowed direction is chosen independently at random among the non-backtracking directions and memory of these choices is lost after one step. Thus
$$
\avg{|\vec{R}(t)|^2} = \sum_{n,n'=0}^{N_M} \avg{\vec{d}_n\cdot \vec{d}_{n'}} \approx
\sum_{n=0}^{N_M} \avg{|\vec{d}_n|^2} +
2\sum_{n=0}^{N_M-1} \avg{\vec{d}_n \cdot \vec{d}_{n+1}},
$$
where all terms $\avg{\vec{d}_n\cdot \vec{d}_{n+m}}$ with $m\neq 0,\pm 1$ have been assumed to be negligible.
For the $\vec{d}_n$ and $\vec{d}_{n+1}$ in the second term there are only 16 possible combinations so this expectation can be computed by enumerating the possible monopole and ice states to obtain
$$
\avg{\vec{d}_n \cdot \vec{d}_{n+1}} = -\frac{a_M^2}{9}.
$$
One thus obtains
$$
\avg{|\vec{R}(t)|^2} \approx N_M a_M^2\left(1-\frac{2}{9}\right) = \frac{7 a_M^2 t}{9 \delta t_M} = 
\frac{14 a^2}{3 \delta t},
$$
taking $N_M \gg 1$. The na\"ive estimate for $D$ is thus reduced by a factor of $7/9$ giving $D \approx {7 a^2}/{6 \delta t}$ or, in natural units $D \approx 7/6 = 1.1\bar{6}$. This estimate can be slightly improved by taking into account the different average densities of type-I and type-II ice states~\cite{nisoli2021color}.

Note that for three-dimensional spin ice this argument follows almost identically. One must only change the dual lattice to diamond lattice with nearest-neighbour distance $a_M = \sqrt{3/2}a_{\rm nn}$ where $a_{\rm nn}$ is the distance between neighboring sites. Given that in three-dimensions 
$\avg{|\vec{R}(t)|^2} \sim 6Dt$, the diffusion constant would be $D \sim 3 a^2/(4 \delta t^2)$ or $D \sim 3/4$ in natural units. The correction due to forbidden hopping directions would give the same reduction of $7/9$, giving $D\sim 7/12 =0.58\bar{3}$ if included.

\section{From Green-Kubo to Monopole Diffusion}
\label{app:green-kubo-diffusion}

In this appendix we show how
the Green-Kubo formula for $\kappa$ can be more
directly related to the monopole diffusion constant by expressing it in terms of fluctuations of the energy polarization
$$
\vec{P}(\vec{\sigma})  \equiv \sum_i \epsilon_i(\vec{\sigma})  \vec{r}_i.
$$
Start by looking at $\vec{J}_t$ (for some given initial state) this naturally says that
$
\vec{J}_t \equiv (\vec{P}_{t+1}-\vec{P}_t)/{\delta t},
$
which can be inverted by summation to yield
$$
\vec{P}_{t}-\vec{P}_0 = \delta t \sum_{\tau=0}^t \vec{J}_{\tau}.
$$
We thus can see that
$$
\avg{(P^\mu_t - P^\mu_0)^2}_{\textrm{eq}} = \delta t^2
\sum^t_{\tau_1=0}\sum^t_{\tau_2=0}
\avg{J^\mu_{\tau_1}J^\mu_{\tau_2}}_{\textrm{eq}}.
$$
Due to symmetry under exchange of $\tau_1$ and $\tau_2$ this sum can be re-written as
$$
\sum^t_{\tau_1=0}\sum^t_{\tau_2=0}
\avg{J^\mu_{\tau_1}J^\mu_{\tau_2}}_{\textrm{eq}} = 
2 \sum^t_{\tau_1=0}\sum^{\tau_1-1}_{\tau_2=0}
\avg{J^\mu_{\tau_1}J^\mu_{\tau_2}}_{\textrm{eq}}
+\sum_{\tau=0}^t 
\avg{J^\mu_{\tau}J^\mu_{\tau}}_{\textrm{eq}},
$$
where the $\tau_1=\tau_2$ case has been handled separately so it is not counted twice. Since there is time translation invariance upon taking the equilibrium average
$
\avg{J^\mu_{\tau_1}J^\mu_{\tau_2}}_{\textrm{eq}} =
\avg{J^\mu_{\tau_1-\tau_2}J^\mu_{0}}_{\textrm{eq}}
$, and therefore
$$
\sum^t_{\tau_1=0}\sum^t_{\tau_2=0}
\avg{J^\mu_{\tau_1}J^\mu_{\tau_2}}_{\textrm{eq}} = 
(t+1)\avg{(J^\mu)^2}_{\textrm{eq}}+
2 \sum^t_{\tau_1=0}\sum^{\tau_1}_{\tau=1}
\avg{J^\mu_{\tau}J^\mu_{0}}_{\textrm{eq}},
$$
where the substitution $\tau \equiv \tau_1-\tau_2$ has been carried out. This can be simplified if the sum over $\tau$ is extended to start at $0$ and compensating terms are added outside, giving
$$
\sum^t_{\tau_1=0}\sum^t_{\tau_2=0}
\avg{J^\mu_{\tau_1}J^\mu_{\tau_2}}_{\textrm{eq}} = 
2 \sum^t_{\tau_1=0}\sum^{\tau_1}_{\tau=0}
\avg{J^\mu_{\tau}J^\mu_{0}}_{\textrm{eq}}
-(t+1)\avg{(J^\mu)^2}_{\textrm{eq}}.
$$
To go further we need to exploit some of the calculus of summations. First define $C_{\tau_1} \equiv \sum^{\tau_1}_{\tau=0} \avg{J^\mu_{\tau}J^\mu_{0}}_{\textrm{eq}}$. The summation by parts identity~\cite{chu2007abel} can be used to write
$$
\sum_{\tau=0}^t C_{\tau} = (t+1)C_{t+1} - \sum^t_{\tau=0} (\tau+1)(C_{\tau+1}-C_{\tau}).
$$
The second piece can be simplified since
$
C_{\tau+1}-C_{\tau} = \avg{J^\mu_{\tau+1}J^\mu_{0}}_{\textrm{eq}}
$.
Putting this together one obtains
\begin{align*}
\sum_{\tau=0}^t C_{\tau} &= 
(t+1)\sum^{t+1}_{\tau=0}
\avg{J^\mu_{\tau}J^\mu_{0}}_{\textrm{eq}} +
\sum^t_{\tau=0} (\tau+1)
\avg{J^\mu_{\tau+1}J^\mu_{0}}_{\textrm{eq}},\\
&=
(t+1)\sum^{t+1}_{\tau=0}
\avg{J^\mu_{\tau}J^\mu_{0}}_{\textrm{eq}} +
\sum^{t-1}_{\tau=1} \tau
\avg{J^\mu_{\tau}J^\mu_{0}}_{\textrm{eq}} .
\end{align*} 
We are mainly interested in the $t\gg 1$ limit where $t\pm 1 \approx t$ so this can be simplified
$$
\sum_{\tau=0}^{t} C_{\tau} 
\approx
t\sum^{\infty}_{\tau=0}
\left(1-\frac{\tau}{t}\right)\avg{J^\mu_{\tau}J^\mu_{0}}_{\textrm{eq}} = 
t\sum^{\infty}_{\tau=0}\avg{J^\mu_{\tau}J^\mu_{0}}_{\textrm{eq}}.
$$
Note that in this last step we have assumed that $\avg{J^\mu_{\tau}J^\mu_{0}}_{\textrm{eq}}$ decays quickly enough that this second term does not survive the large $t$ limit~\footnote{A correlation function $\avg{J^\mu_{\tau}J^\mu_{0}}_{\textrm{eq}}$ that decays like a power-law (or does not decay at all), as may happen for ballistic transport, would violate this assumption.}
Putting this all back together (for $t\gg 1$)
\begin{align*}
\avg{(P^\mu_t - P^\mu_0)^2}_{\textrm{eq}} &\approx \delta t^2\left(
2 \sum^{t}_{\tau=0}C_{\tau}
-t\avg{(J^\mu)^2}_{\textrm{eq}}
\right), \\
&\approx
2t \delta t^2
\sum^{\infty}_{\tau=0}\avg{J^\mu_{\tau}J^\mu_{0}}_{\textrm{eq}}\left(1-\frac{1}{2}\delta_{\tau,0}\right).
\end{align*}
The diagonal parts of the thermal conductivity can then be written cleanly as
$$
\kappa_{\mu\mu} = \frac{1}{\delta t k_{\rm B} V T^2}\left\{
\lim_{t \rightarrow \infty} \frac{\avg{(P^\mu_t - P^\mu_0)^2}_{\textrm{eq}}}{2t}
\right\}.
$$
The expression in the brackets is similar to a the definition of the diffusion constant for a particle, except instead of involving the monopole position $\vec{R}$ it involves the energy polarization $\vec{P}$.

In the low temperature limit of square ice $\vec{P}$ can be related to the monopole positions $\vec{R}$. First, shift the energy so the ice states have energy zero and the local energy density is
$$
\epsilon_i = \frac{J}{2}\sigma_i\left(\sum_j \sigma_j\right) + J = J\sigma_i(-1)^I (Q_I-Q_J),
$$
where $I$ and $J$ are the two neighbouring tetrahedra associated with site $i$ and $Q_I \equiv \frac{1}{2} (-1)^I \sum_{i \in I} \sigma_i$. Thus any site which belongs to two ice tetrahedra has $\epsilon_i=0$. If the defects (tetrahedra with $Q_I \neq 0$) are dilute then it can be safely assumed that (with high probability) a site only belongs to one, thus
$$
\vec{P} \approx \sum_{Q_I \neq 0} \sum_{i \in I} \epsilon_i \vec{r}_i.
$$
In the unlikely event two defects are neighbours then this expression counts them twice, requiring a correction term. For a tetrahedron with a single monopole this yields
$$
 \sum_{i \in I} \epsilon_i \vec{r}_i = 2J(\vec{R}-\vec{d}),
$$
where $\vec{R}$ is the center of the tetrahedron and $\vec{d}$ is the ``polarization'' vector pointing from the tetrahedron center to the minority spin. Thus with $N_1$ monopoles at locations $\vec{R}_n$ with polarizations $\vec{d}_n$ then
$$
\vec{P} \approx 2J\sum_{n=1}^{N_1} (\vec{R}_n-\vec{d}_n).
$$
If monopole motion is assumed to be approximately uncorrelated, then only terms with $n=n'$ need to kept and and thus
$$
\avg{|\vec{P}_t - \vec{P}_0|^2}_{\textrm{eq}}
\approx (2J)^2 N_1
\avg{|\vec{R}(t)-\vec{R}(0)-\vec{d}(t)+\vec{d}(0)|^2}_{\textrm{eq}},
$$
under the assumption all monopoles are identical and thus the sum simply gives the total number of monopoles $N_1$ times the correlator for a single monopole.

If at long times $\vec{R}$ and $\vec{d}$ are uncorrelated, $\vec{d}$ is unimportant. Consider the quantity
$\avg{|\vec{R}(t)-\vec{R}(0)-\vec{d}(t)+\vec{d}(0)|^2}_{\textrm{eq}}$ which can be written
$$
\avg{|\vec{R}(t)-\vec{R}(0)|^2}_{\textrm{eq}}
+\avg{|\vec{d}(t)-\vec{d}(0)|^2}_{\textrm{eq}}
+4\avg{(\vec{R}(t)-\vec{R}(0))\cdot\vec{d}(0)}_{\textrm{eq}},
$$
where we have assumed time-reversal symmetry to set 
$\avg{\vec{R}(t)\cdot\vec{d}(0)}_{\textrm{eq}}=\avg{\vec{R}(0)\cdot\vec{d}(t)}_{\textrm{eq}}$. The last term vanishes since $\avg{\vec{R}}_{\textrm{eq}}=\avg{\vec{d}}_{\textrm{eq}}=0$ absent correlations in $\vec{R}$ and $\vec{d}$~\footnote{For diffusive behaviour we expect that $|\vec{R}(t)| \sim O(\sqrt{t})$ and so this term may vanish even if $\vec{R}$ and $\vec{d}$ are correlated, once the $t\rightarrow\infty$ limit is taken.}. The second term does not survive the $t\rightarrow \infty$ limit as it is bounded, with $\avg{|\vec{d}(t)-\vec{d}(0)|^2}_{\textrm{eq}} \leq 2$ (since $|\vec{d}|=1/\sqrt{2}$). Thus we have
$$
\lim_{t\rightarrow \infty}\left\{ \frac{\avg{|\vec{P}_t - \vec{P}_0|^2}_{\textrm{eq}}}{4t \delta t}\right\}
\approx
(2J)^2 N_1
\lim_{t\rightarrow \infty} \left\{
\frac{\avg{|\vec{R}(t)-\vec{R}(0)|^2}_{\textrm{eq}}}{4t \delta t}
\right\}
$$
This quantity in terms of $\vec{P}$ has thus been related to the monopole diffusion constant
$$
D \equiv \lim_{t\rightarrow \infty} \left\{
\frac{\avg{|\vec{R}(t)-\vec{R}(0)|^2}_{\textrm{eq}}}{4t \delta t}
\right\}.
$$
For the isotropic thermal conductivity $\kappa \equiv (\kappa_{xx}+\kappa_{yy})/2$ we then have
$$
\kappa = 
 \frac{1}{ k_{\rm B} V T^2}\left\{
\lim_{t \rightarrow \infty} \frac{\avg{|\vec{P}_t - \vec{P}_0|^2}_{\textrm{eq}}}{4t\delta t}
\right\}=
\frac{(2J)^2 N_1}{k_{\rm B} V T^2} D.
$$
For a dilute gas of monopoles one has $E \sim 2J N_1$ where $N_1 \approx N e^{-2J/(k_{\rm B} T)}$. This gives the heat capacity per unit volume of $C \sim k_{\rm B} [2J/(k_{\rm B} T)]^2 (N_1/V)$ and thus
$D \approx \kappa/C$.

\bibliography{draft}
\end{document}